\documentclass[lettersize,journal]{IEEEtran}
\usepackage{amsmath,amsfonts}
\usepackage{algorithmic}
\usepackage{algorithm}
\usepackage{array}
\usepackage{textcomp}
\usepackage{stfloats}
\usepackage{url}
\usepackage{verbatim}
\usepackage{graphicx}
\usepackage{cite}
\usepackage{diagbox}
\usepackage{indentfirst}
\usepackage{multirow}
\usepackage{graphicx}
\usepackage{color}
\usepackage{xcolor}
\usepackage{colortbl}
\usepackage{listings}
\usepackage{enumerate}
\usepackage{amsmath,amsthm,amscd,amssymb,latexsym,upref,stmaryrd}
\usepackage{amsfonts,epsfig}
\usepackage{CJK}
\usepackage{float}
\usepackage{cite}
\usepackage{lipsum}
\usepackage{multicol}
\usepackage{setspace}
\usepackage{bm}
\usepackage{amsthm}
\usepackage{amsmath}
\usepackage{amssymb}
\usepackage{amsfonts}
\usepackage{arydshln}
\usepackage{stfloats}
\usepackage{booktabs}
\usepackage{threeparttable}
\usepackage{stackengine}
\newtheorem{remark}{Remark}
\usepackage[utf8]{inputenc}
\usepackage{fancyhdr}

\usepackage[colorlinks,linkcolor=purple,anchorcolor=purple,citecolor=purple]{hyperref}

\graphicspath{{fig/}}
\usepackage{makecell}

\newcommand{\xrowht}[2][0]{\addstackgap[.5\dimexpr#2\relax]{\vphantom{#1}}}

\usepackage{subcaption}

\newcommand{\bl}[1]{\textcolor{blue}{#1}}

\DeclareMathOperator*{\argmin}{arg\,min}

\begin{document}

\title{Learned Off-Grid Imager for Low-Altitude Economy with Cooperative ISAC Network}

\author{Yixuan~Huang, Jie~Yang, Shuqiang Xia, Chao-Kai~Wen,~\IEEEmembership{Fellow,~IEEE,} and Shi~Jin,~\IEEEmembership{Fellow,~IEEE\vspace{-0.5cm}}
\thanks{Yixuan~Huang and Shi~Jin are with the National Mobile Communications Research Laboratory, Southeast University, Nanjing 210096, China (e-mail: \{huangyx; jinshi\}@seu.edu.cn).
Jie~Yang is with the Key Laboratory of Measurement and Control of Complex Systems of Engineering, Ministry of Education, Southeast University, Nanjing 210096, China (e-mail: yangjie@seu.edu.cn).
Jie~Yang and Shi~Jin are also with the Frontiers Science Center for Mobile Information Communication and Security, Southeast University, Nanjing 210096, China.
Chao-Kai Wen is with the Institute of Communications Engineering, National Sun Yat-sen University, Kaohsiung 80424, Taiwan. (e-mail: chaokai.wen@mail.nsysu.edu.tw).
Shuqiang Xia is with ZTE Corporation and the State Key Laboratory of Mobile Network and Mobile Multimedia Technology, Shenzhen 518055, China (e-mail: xia.shuqiang@zte.com.cn).
Part of this paper will be presented in VTC2025-Spring \cite{huang2025cooperative}.}
}

\maketitle

\begin{abstract}

The low-altitude economy is emerging as a key driver of future economic growth, necessitating effective flight activity surveillance using existing mobile cellular network sensing capabilities. However, traditional monostatic and localization-based sensing methods face challenges in fusing sensing results and matching channel parameters.
To address these challenges, we model low-altitude surveillance as a compressed sensing (CS)-based imaging problem by leveraging the cooperation of multiple base stations and the inherent sparsity of aerial images. Additionally, we derive the point spread function to analyze the influences of different antenna, subcarrier, and resolution settings on the imaging performance.
Given the random spatial distribution of unmanned aerial vehicles (UAVs), we propose a physics-embedded learning method to mitigate off-grid errors in traditional CS-based approaches. Furthermore, to enhance rare UAV detection in vast low-altitude airspace, we integrate an online hard example mining scheme into the loss function design, enabling the network to adaptively focus on samples with significant discrepancies from the ground truth during training.
Simulation results demonstrate the effectiveness of the proposed low-altitude surveillance framework. The proposed physics-embedded learning algorithm achieves a 97.55\% detection rate, significantly outperforming traditional CS-based methods under off-grid conditions.
Part of the source code for this paper will be soon accessed at \url{https://github.com/kiwi1944/LAEImager}.
\end{abstract}


\section{Introduction}
The low-altitude economy (LAE) has grown rapidly, enabling applications such as food delivery, traffic monitoring, and agricultural irrigation \cite{wu2024vehicle,zheng2024random,he2024device}. Currently, LAE primarily operates in uncontrolled airspace below 300 meters \cite{ChinaMobile2024WhitePaper}. The surge in unmanned aerial vehicle (UAV) deployments necessitates advanced surveillance techniques to ensure flight safety. According to 3GPP specifications \cite{ChinaMobile2024WhitePaper,3gpp2023study}, UAV localization with meter-level accuracy is essential for intrusion detection and trajectory planning. However, UAVs’ low radar cross-section (RCS) and high mobility pose significant challenges for real-time sensing algorithm design \cite{wu2024vehicle}. This study aims to develop all-weather, around-the-clock sensing techniques for aerial surveillance.

Conventional UAV localization relies on the global navigation satellite system (GNSS), which is vulnerable to blockage and jamming, often resulting in positioning failures \cite{ruan2021cooperative,meles2023performance}. Unauthorized UAVs may also disable GNSS receivers or reject localization signals, further complicating surveillance.  
Visible-light cameras have been integrated with radio frequency (RF) sensing via attention-based spatio-temporal fusion techniques \cite{wu2024vehicle}. However, their performance degrades under low-visibility conditions, such as nighttime or haze, and they struggle with long-range target detection \cite{zhao2022vision}. Moreover, these approaches require additional sensing hardware, increasing deployment costs for large-scale networks.  
Integrated sensing and communication (ISAC) provides a cost-effective alternative by leveraging existing cellular infrastructure without the need for extra sensors or hardware \cite{huang2025integrated,liu2024cooperative,li2023towards,cheng2024networked}.

\textit{Related Work}---UAV sensing using cellular networks can be categorized into active and passive paradigms \cite{wu2024vehicle}.

In the active paradigm, UAVs act as cooperative devices that establish communication links with base stations (BSs). By utilizing pilot transmissions and channel estimation, delay and angular parameters can be extracted and mapped to UAV locations using geometric relationships \cite{meles2023performance}.  
Deep supervised learning and reinforcement learning techniques have been applied to achieve 3D UAV localization based on received signal strength measurements from surrounding BSs \cite{afifi2021autonomous}.
Although conventional user localization algorithms can be adapted for UAV tracking \cite{he2024device,huang2023joint}, they are ineffective for monitoring uncooperative UAVs.

The passive sensing paradigm addresses this limitation by eliminating the need for target cooperation.  
Analogous to monostatic radar systems, beamforming and scanning can generate ``range-angle'' maps of the airspace \cite{guan20213,tang2024cooperative}, but require large antenna arrays to ensure high sensing accuracy.
Graph neural networks have been adopted to design ISAC beamformers for enhanced sensing \cite{wang2024heterogeneous}, while UAV location features can be extracted using convolutional neural networks (CNNs) \cite{wang2021deep}.
However, beam scanning is time-consuming \cite{ma2024networked}, and fusing multi-BS sensing data involves complex decision-level algorithms \cite{tang2024cooperative}.

To improve sensing efficiency, bistatic and multi-static configurations enable BSs to jointly transmit and receive sensing signals \cite{liu2024cooperative,wang2024heterogeneous}. Most passive localization approaches estimate delay and angular parameters before mapping them to location coordinates \cite{liu2024cooperative,huang2023joint,ma2024networked}. This two-step process, however, is vulnerable to error propagation, where estimation inaccuracies degrade localization performance \cite{shih2024machine}.  
Furthermore, in multi-target scenarios, data association becomes challenging, requiring complex matching of multi-dimensional channel parameters to the corresponding UAVs \cite{shi2022device,shi2024joint}.

\textit{Proposed Approach}—Given the relatively small UAV size compared to the vast low-altitude 3D space, we formulate UAV surveillance as a compressed sensing (CS)-based passive imaging problem by discretizing the aerial space into grids, fundamentally differing from previous studies.  
Specifically, ``{\it imaging}'' here refers to obtaining the scattering coefficient image of the aerial space, capturing the existence, location, scattering, and swarm shape information of low-altitude targets.
Using CS-based imaging algorithms \cite{dai2009subspace,tong2021joint,huang2024ris}, raw channel state information (CSI) measurements are directly processed to generate images, enabling multi-static cooperation while mitigating error propagation and data association challenges.
Despite these advantages, three major challenges remain: system configuration, on-grid approximation errors, and the high sparsity of low-altitude images. 

{\it First}, traditional cellular networks are designed for ground communication and must be adapted for low-altitude sensing. The required hardware configuration to ensure high sensing performance remains unclear.  
In this study, we introduce the point spread function (PSF), a noise-free and target-independent metric derived from the sensing matrix, to evaluate aerial imaging performance and analyze the impact of system parameters.  
PSF analysis guides ISAC system configuration and algorithm design, including antenna array layout, subcarrier and bandwidth selection, and imaging region and resolution. 

{\it Second}, traditional CS models assume that targets lie exactly on predefined grid points \cite{dai2009subspace}, while UAV trajectories are inherently continuous, leading to off-grid modeling errors that distort the imaging results \cite{jing2024isac}.  
A common solution is to generate an initial estimate using on-grid models and refine it iteratively toward off-grid locations \cite{shang2024mixed,you2022bayesian}. However, this approach relies heavily on the accuracy of the initial estimate \cite{ma2024networked}. 
Alternatively, off-grid deviations can be modeled as unknown parameters embedded in the sensing matrix and estimated jointly with the sparse vector \cite{yang2012off,wei2022off}, though this method involves complex Taylor expansion and derivative computation in Cartesian coordinates.
Gridless methods such as atomic norm minimization (ANM) address off-grid errors through convex optimization in continuous domains \cite{li2024atomic,gao2024robust}, but incur heavy computational costs and memory usage, especially in high-dimensional settings.

To address these limitations, deep learning (DL) has been applied in off-grid CS problems, particularly for channel estimation \cite{huang2023off,zhang2024two}, where deep neural networks (DNNs) learn the mapping between CSI and sparse representations.  
Although DNN-based techniques have been developed for low-altitude surveillance \cite{wu2024vehicle,afifi2021autonomous,wang2024heterogeneous,wang2021deep}, black-box DNNs suffer from limited interpretability and high sensitivity to training data. Thus, researchers have incorporated physical models into learning, categorized into three approaches \cite{guo2023physics,chen2020review}:
1) \textbf{Learning after physics processing} uses on-grid models for initial outputs, which are refined by DNNs \cite{wu2019deep};  
2) \textbf{Learning with physics loss} embeds physical models into the loss function, but remains sensitive to off-grid mismatches;  
3) \textbf{Learning with physics models} unfolds model-based CS algorithms into DNN layers, but suffers from high complexity due to intricate model expressions \cite{wan2021deep,su2023real}.

Building on these insights, we propose a \textbf{learned physics-embedded off-grid imager} for UAV surveillance. This hybrid framework first applies on-grid models for coarse results and then refines them using DNNs to mitigate off-grid errors while maintaining computational efficiency. 

{\it Finally}, extreme sparsity in low-altitude images poses a major challenge. Unlike target images like those in MNIST dataset, where structured features aid in object detection, low-altitude images consist mostly of zero-valued voxels, with only a few non-zero points corresponding to UAVs. This severe imbalance between zero and non-zero voxels poses challenges for neural network training.
Previous studies in channel estimation fields typically employ the mean square error (MSE) loss function for DNN training \cite{huang2023off,zhang2024two,wu2019deep,wan2021deep,su2023real}. However, it may not be effective in low-altitude imaging scenarios. Cross-entropy loss, used in \cite{hu2021metasensing,wang2024dreamer}, detects target presence and generates binary (``0-1'') images but fails to preserve scattering coefficient information, which is essential for target characterization.

To overcome these limitations, we adopt online hard example mining (OHEM) \cite{shrivastava2016training}, a computer vision technique prioritizing samples with large discrepancies from the ground truth, to develop tailored loss functions. This approach effectively optimizes the DNN to simultaneously enhance target detection and preserve scattering information, achieving high detection rates (DRs) and low false alarm rates (FARs).

In summary, this study makes the following contributions:
\begin{itemize}
\item \textbf{Cooperative Low-Altitude Imaging:} We reformulate aerial surveillance as a CS-based imaging problem through space discretization. Our approach fully exploits the sensing capabilities of cooperative ISAC networks, mitigating error propagation and data association challenges. We also derive the PSF to analyze the impact of system parameters and qualitatively evaluate imaging capability.

\item \textbf{Learned Physics-Embedded Off-Grid Imager:} We address off-grid challenges by integrating physical and data-driven models. Our method first applies on-grid models, then refines the results using DNNs. We also propose novel OHEM-based loss functions tailored for low-altitude imaging. Simulation results demonstrate significant improvements over conventional methods.
\end{itemize}

The remainder of this paper is structured as follows:
Sec. \ref{sec-system-model} introduces the ISAC system model.
Sec. \ref{sec-problem-formulation} formulates the imaging-based low-altitude surveillance problem.
Sec. \ref{sec-imaging-algorithm} presents the proposed imaging algorithms, particularly for off-grid scenarios.
Sec. \ref{sec-simu} provides extensive simulation results.
Sec. \ref{sec-conclusion} concludes this paper.

\textit{Notations}---Scalars (e.g., $a$) are denoted in italics, vectors (e.g., $\mathbf{a}$) in bold lowercase, and matrices (e.g., $\mathbf{A}$) in bold uppercase. The modulus of $a$ is represented as $|a|$, and the imaginary unit is denoted by $j = \sqrt{-1}$. The $\ell_{1}$- and $\ell_{2}$-norm of $\mathbf{a}$ are given by $\|\mathbf{a}\|_{1}$ and $\|\mathbf{a}\|_{2}$, respectively. The notation $\text{diag}{(\mathbf{a})}$ constructs a diagonal matrix with the elements of $\mathbf{a}$. The transpose and Hermitian (conjugate transpose) operators are denoted by $(\cdot)^{\text{T}}$ and $(\cdot)^{\text{H}}$, respectively. The inner product of two vectors $\mathbf{a}$ and $\mathbf{b}$ is represented by $\left<\mathbf{a}, \mathbf{b}\right>$.

\begin{table*}[t]
	\centering
	\caption{Notations of important variables.}\label{tab-notation}
	\renewcommand{\arraystretch}{1.6}
	\fontsize{8}{8}\selectfont
	\begin{tabular}{m{1.3cm} m{6.8cm} m{1.3cm} m{6.8cm}}
		\specialrule{1pt}{0pt}{-1pt}
        \xrowht{10pt}
		Notation & Definition & Notation & Definition \\
		\hline
		$N_{\text{b}}$ & number of BSs & $N_0$ & number of antennas in BS UPAs along one dimension \\
		$f_0$ & center carrier frequency & $\lambda_0$ & center carrier wavelength \\
		$K$ & number of communication users & $\xi$ & antenna spacing scale for sparse arrays \\
		$B$ & signal bandwidth & $N_{\text{f}}$ & number of subcarriers \\
		$\mathbf{w}_{n_{\text{b}}, n_0}$ & beamforming vector of the $n_0$-th RF chain at the $n_{\text{b}}$-th BS & $\mathbf{W}_{n_{\text{b}}}$ & $\mathbf{W}_{n_{\text{b}}} = [\mathbf{w}_{n_{\text{b}}, 1}, \ldots, \mathbf{w}_{n_{\text{b}}, N_0^2}]$, beamforming matrix \\
		${\mathbf{x}}_{n_{\text{b}}, n_{\text{f}}, n_{\text{s}}}^{\text{c}}$ & normalized communication signal at the $n_{\text{b}}$-th BS, the $n_{\text{f}}$-th subcarrier, and the $n_{\text{s}}$-th symbol interval & ${\mathbf{x}}_{n_{\text{b}}, n_{\text{f}}, n_{\text{s}}}^{\text{s}}$ & normalized sensing signal at the $n_{\text{b}}$-th BS, the $n_{\text{f}}$-th subcarrier, and the $n_{\text{s}}$-th symbol interval \\
		${\mathbf{x}}_{n_{\text{b}}, n_{\text{f}}, n_{\text{s}}}$ & ISAC signal at the $n_{\text{b}}$-th BS, the $n_{\text{f}}$-th subcarrier, and the $n_{\text{s}}$-th symbol interval & $P_{\text{t}}$ & total transmit power of one BS \\
		$\mathbf{h}_{n_{\text{b}}, k, n_{\text{f}}}$ & channel between the $n_{\text{b}}$-th BS and the $k$-th user on the $n_{\text{f}}$-th subcarrier & $\tilde{{z}}^{\text{c}}_{k, n_{\text{f}},n_{\text{s}}}$ & AWGN at the $k$-th user, the $n_{\text{f}}$-th subcarrier, and the $n_{\text{s}}$ symbol interval \\
		$\zeta^2_{\text{c}}$ & variance of $\tilde{{z}}^{\text{c}}_{k, n_{\text{f}},n_{\text{s}}}$ & $\gamma_{k, n_{\text{f}}}$ & SINR of the $k$-th user on the $n_{\text{f}}$-th subcarrier \\
		$\mathbf{h}_{k, n_{\text{f}}}$ & $\mathbf{h}_{k, n_{\text{f}}} = [\mathbf{h}^{\text{T}}_{1, k, n_{\text{f}}}, \ldots, \mathbf{h}^{\text{T}}_{N_{\text{b}}, k, n_{\text{f}}}]^{\text{T}}$ & $\mathbf{w}_{k}$ & $\mathbf{w}_{k} = [\mathbf{w}^{\text{T}}_{1, k}, \ldots, \mathbf{w}^{\text{T}}_{N_{\text{b}}, k}]^{\text{T}}$ \\
		$\mathbf{r}^{\text{s}}_{n_{\text{b}1},n_{\text{b}2},n_{\text{f}},n_{\text{s}}}$ & received sensing signal at the $n_{\text{b2}}$-th BS, transmitted by the $n_{\text{b1}}$-th BS at the $n_{\text{f}}$-th subcarrier and the $n_{\text{s}}$ symbol interval & $\mathbf{H}_{n_{\text{b}1},n_{\text{b}2},n_{\text{f}}}$ & channel from the $n_{\text{b}1}$-th BS, scattered by the ROI, and received at the $n_{\text{b2}}$-th BS \\
		$\mathbf{F}_{n_{\text{b}2}}$ & combiner at the $n_{\text{b2}}$-th BS & $\tilde{\mathbf{z}}_{n_{\text{b}1},n_{\text{b}2},n_{\text{f}},n_{\text{s}}}^{\text{s}}$ & additive noise in $\mathbf{r}^{\text{s}}_{n_{\text{b}1},n_{\text{b}2},n_{\text{f}},n_{\text{s}}}$ \\
		$h^{\text{s}}_{n_{\text{t}},n_{\text{r}},n_{\text{f}}}$ & $(n_{\text{t}},n_{\text{r}})$-th element in $\mathbf{H}_{n_{\text{b}1},n_{\text{b}2},n_{\text{f}}}$ & $G_{\text{s}}$ & combined antenna gain at the transmitter and receiver \\
		$\breve{\sigma}(x, y, z)$ & continuous ROI image & $\lambda_{n_{\text{f}}}$ & wavelength of the $n_{\text{f}}$-th subcarrier \\
		$d_1(x,y,z)$ & distances from $[x, y, z]^{\text{T}}$ to the $n_{\text{t}}$-th transmitting antenna & $d_2(x,y,z)$ & distances from $[x, y, z]^{\text{T}}$ to the $n_{\text{r}}$-th receiving antenna \\
		$N_{\text{s}}$ & $N_{\text{s}}=N_0^2$, number of symbol intervals & $\mathbf{R}^{\text{s}}_{n_{\text{b}1},n_{\text{b}2},n_{\text{f}}}$ & $\mathbf{R}^{\text{s}}_{n_{\text{b}1},n_{\text{b}2},n_{\text{f}}} = [\mathbf{r}^{\text{s}}_{n_{\text{b}1},n_{\text{b}2},n_{\text{f}},1}, \ldots, \mathbf{r}^{\text{s}}_{n_{\text{b}1},n_{\text{b}2},n_{\text{f}},N_{\text{s}}}]$ \\
		$\mathbf{X}_{n_{\text{b}1},n_{\text{f}}}$ & $\mathbf{X}_{n_{\text{b}1},n_{\text{f}}}=[\mathbf{x}_{n_{\text{b1}}, n_{\text{f}}, 1}, \ldots, \mathbf{x}_{n_{\text{b1}}, n_{\text{f}}, N_{\text{s}}}]$ & $\tilde{\mathbf{Z}}^{\text{s}}_{n_{\text{b}1},n_{\text{b}2},n_{\text{f}}}$ & $\tilde{\mathbf{Z}}^{\text{s}}_{n_{\text{b}1},n_{\text{b}2},n_{\text{f}}}=[\tilde{\mathbf{z}}^{\text{s}}_{n_{\text{b}1},n_{\text{b}2},n_{\text{f}},1}, \ldots, \tilde{\mathbf{z}}^{\text{s}}_{n_{\text{b}1},n_{\text{b}2},n_{\text{f}},N_{\text{s}}}]$ \\
		$\widehat{\mathbf{H}}_{n_{\text{b}1},n_{\text{b}2},n_{\text{f}}}$ & estimate of ${\mathbf{H}}_{n_{\text{b}1},n_{\text{b}2},n_{\text{f}}}$ & $y_{n_{\text{t}},n_{\text{r}},n_{\text{f}}}$ & $(n_{\text{t}},n_{\text{r}})$-th element of $\widehat{\mathbf{H}}_{n_{\text{b}1},n_{\text{b}2},n_{\text{f}}}$ \\
		$N_{\text{v}}$ & number of voxels in the ROI & $\boldsymbol{\sigma}$ & $\boldsymbol{\sigma} = [\sigma_1, \ldots, \sigma_{N_{\text{v}}}]^{\text{T}}$, discretized version of $\breve{\sigma}(x, y, z)$ \\
		$\sigma_{n_{\text{v}}}$ & scattering coefficient of the $n_{\text{v}}$-th voxel & $M$ & number of UAVs in the ROI \\
		$\mathbf{y}$ & CSI measurements of all BSs & $\mathbf{A}$ & sensing matrix related to all BSs \\
		$\mathbf{z}$ & AWGN involved in $\mathbf{y}$ & $\zeta^2_{\text{s}}$ & variance of $\mathbf{z}$ \\
		$\hat{\boldsymbol{\sigma}}$ & estimate of $\boldsymbol{\sigma}$ & $\varepsilon$ & reconstruction accuracy threshold \\
		\specialrule{1pt}{0pt}{0pt}
	\end{tabular}
\end{table*}

\section{System Model}
\label{sec-system-model}

\begin{figure}
    \centering
    \includegraphics[width=0.88\linewidth]{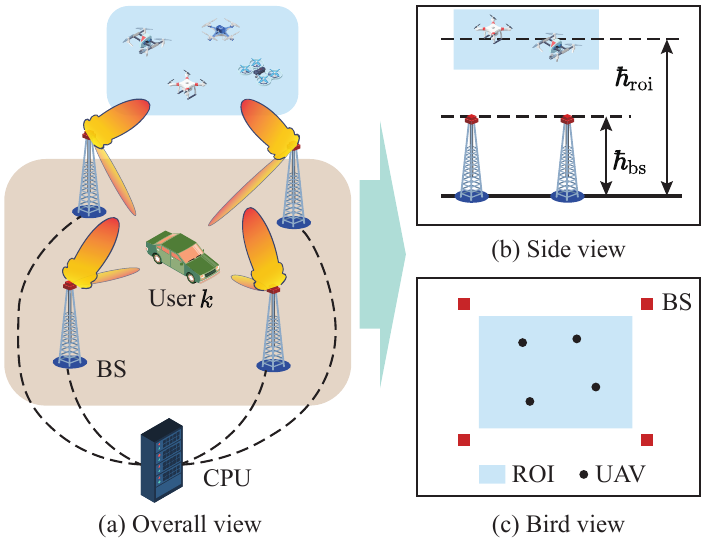}
    \captionsetup{font=footnotesize}
    \caption{Illustration of the cooperative ISAC network.}
    \label{fig-model}
\end{figure}

We consider an ISAC system operating within a 3D space, represented as $[x,y,z]^{\text{T}} \in \mathbb{R}^3$, as illustrated in Fig.~\ref{fig-model}.
The system comprises $N_{\text{b}}$ BSs deployed at an altitude of $\hbar_{\text{bs}}$, forming a convex region with $N_{\text{b}}$ edges in the horizontal plane.
Each BS is equipped with a full-duplex uniform planar array (UPA) consisting of $N_0\times N_0$ antennas.
The UPAs are vertically aligned to the ground, with their normal vectors directed toward the convex region's center. The antenna spacing is given by $(\lambda_0/2) \cdot \xi $, where $\lambda_0$ denotes the wavelength corresponding to the center carrier frequency $f_0$.
To enhance spatial resolution, we may consider sparse antenna arrays \cite{lu2024tutorial,chen2024near}, where $\xi\ge 1$.
Self-interference at the full-duplex BSs is mitigated through antenna separation and optimized beamforming \cite{zhang2015full}.
The $N_{\text{b}}$ BSs are synchronized via optical fiber to ensure precise timing. The region of interest (ROI), depicted in Fig.~\ref{fig-model}, represents a large surveillance area at altitude $\hbar_{\text{roi}}$.
Its boundaries are predefined based on sensing requirements \cite{li2023towards}.
Specifically, the BSs are properly mounted and selected to ensure a line-of-sight (LOS) path between them and the ROI.
For signal transmission, the system employs orthogonal frequency division multiplexing (OFDM) with $N_{\text{f}}$ subcarriers and a total bandwidth of $B$.
The key variables in the system model are summarized in Table \ref{tab-notation}.

\subsection{Transmit Signal Model}

We consider a scenario where the BSs simultaneously sense low-altitude flight activities while communicating with $K$ single-antenna users in the downlink mode \cite{li2023towards,wang2024heterogeneous}. To achieve this, the BSs form a wide beam to cover the ROI.
At the $n_{\text{b}}$-th BS, the $n_{\text{f}}$-th subcarrier, and the $n_{\text{s}}$-th symbol interval, the communication and sensing signals are designed to be spatially orthogonal, given by \cite{li2023towards}:
\begin{subequations}
\begin{align}
\bar{\mathbf{x}}_{n_{\text{b}}, n_{\text{f}}, n_{\text{s}}}^{\text{c}} &= \sum_{k = 1}^K\mathbf{w}_{n_{\text{b}}, k} s_{k, n_{\text{f}}, n_{\text{s}}}, \\
\bar{\mathbf{x}}_{n_{\text{b}}, n_{\text{f}}, n_{\text{s}}}^{\text{s}} &= \sum_{n_0 = K+1}^{N_0^2} \mathbf{w}_{n_{\text{b}}, n_0} s_{n_{\text{b}}, n_0, n_{\text{f}}, n_{\text{s}}},
\end{align}
\end{subequations}
where $s_{k, n_{\text{f}}, n_{\text{s}}}$ is the information-bearing data for the $k$-th user, and $s_{n_{\text{b}}, n_0, n_{\text{f}}, n_{\text{s}}}$ represents the dedicated sensing signal processed by the $n_0$-th RF chain of the $n_{\text{b}}$-th BS.
The beamforming matrix is defined as $\mathbf{W}_{n_{\text{b}}} = [\mathbf{w}_{n_{\text{b}}, 1}, \ldots, \mathbf{w}_{n_{\text{b}}, N_0^2}]\in\mathbb{C}^{N_0^2\times N_0^2}$ and can be designed based on the approaches proposed in \cite{li2023towards,wang2024heterogeneous}.
Consequently, the transmitted ISAC signal at the $n_{\text{b}}$-th BS can be expressed as
\begin{equation}
\mathbf{x}_{n_{\text{b}}, n_{\text{f}}, n_{\text{s}}} = \sqrt{P_{\text{c}}}\mathbf{x}_{n_{\text{b}}, n_{\text{f}}, n_{\text{s}}}^{\text{c}} + \sqrt{P_{\text{s}}}\mathbf{x}_{n_{\text{b}}, n_{\text{f}}, n_{\text{s}}}^{\text{s}},
\end{equation}
where $\mathbf{x}_{n_{\text{b}}, n_{\text{f}}, n_{\text{s}}}^{\text{c}}$ and $\mathbf{x}_{n_{\text{b}}, n_{\text{f}}, n_{\text{s}}}^{\text{s}}$ are the normalized versions of $\bar{\mathbf{x}}_{n_{\text{b}}, n_{\text{f}}, n_{\text{s}}}^{\text{c}}$ and $\bar{\mathbf{x}}_{n_{\text{b}}, n_{\text{f}}, n_{\text{s}}}^{\text{s}}$, respectively. Here,
$P_{\text{c}}$ and $P_{\text{s}}$ represent the power allocated to communication and sensing, respectively.
The total transmit power at the $n_{\text{b}}$-th BS is $P_{\text{t}} = P_{\text{c}} + P_{\text{s}}$.
For simplicity, we assume that $P_{\text{t}}$, $P_{\text{c}}$, and $P_{\text{s}}$ are identical across all BSs, and that the signals $\mathbf{x}_{n_{\text{b}}, n_{\text{f}}, n_{\text{s}}}$ transmitted by different BSs are orthogonal \cite{liu2024cooperative,tang2024cooperative}.
Furthermore, specially designed protocols can assist in suppressing collaborative sensing interference \cite{ma2024networked}.

\subsection{Communication Model}

The wireless channel between the $n_{\text{b}}$-th BS and the $k$-th user on the $n_{\text{f}}$-th subcarrier is denoted as $\mathbf{h}_{n_{\text{b}}, k, n_{\text{f}}}^{\text{H}}$, which includes both direct BS-user multipaths and BS-ROI-user interference paths. Given the transmitted signal $\mathbf{x}_{n_{\text{b}}, n_{\text{f}}, n_{\text{s}}}$, the received signal at the $k$-th user is expressed in \eqref{eq-commun-signal-model} (at the top of this page) \cite{li2023towards,wang2024heterogeneous}, where $\tilde{{z}}^{\text{c}}_{k, n_{\text{f}},n_{\text{s}}}$ represents additive white Gaussian noise (AWGN) with variance $\zeta^2_{\text{c}}$.
\begin{figure*}[t]
\begin{equation}
\label{eq-commun-signal-model}
{r}^{\text{c}}_{k, n_{\text{f}}, n_{\text{s}}}=
\!\underbrace{\sum_{n_{\text{b}}=1}^{N_{\text{b}}} \mathbf{h}_{n_{\text{b}}, k, n_{\text{f}}}^{\text{H}}\mathbf{w}_{n_{\text{b}}, k}s_{k, n_{\text{f}}, n_{\text{s}}}}_{\text{Desired communication signal}} + 
\!\underbrace{\sum_{n_{\text{b}}=1}^{N_{\text{b}}} \sum_{\substack{i=1 \\ i\neq k}}^K \mathbf{h}_{n_{\text{b}}, k, n_{\text{f}}}^{\text{H}}\mathbf{w}_{n_{\text{b}}, i}s_{i, n_{\text{f}}, n_{\text{s}}}}_{\text{Multi-user interference}} + 
\!\underbrace{\sum_{n_{\text{b}}=1}^{N_{\text{b}}}\sum_{n_0 = K+1}^{N_0^2} \mathbf{h}_{n_{\text{b}}, k, n_{\text{f}}}^{\text{H}} \mathbf{w}_{n_{\text{b}}, n_0} s_{n_{\text{b}}, n_0, n_{\text{f}}, n_{\text{s}}}}_{\text{Sensing interference}} + 
\underbrace{\tilde{{z}}^{\text{c}}_{k, n_{\text{f}}, n_{\text{s}}}}_{\text{Noise}},
\end{equation}
\hrulefill
\end{figure*}
The signal-to-interference-plus-noise ratio (SINR) for the $k$-th user on the $n_{\text{f}}$-th subcarrier is given by 
\begin{equation}\label{eq-user-sinr}
\gamma_{k, n_{\text{f}}} = \frac{|\mathbf{h}_{k, n_{\text{f}}}^{\text{H}}\mathbf{w}_{k}|^2}{\sum_{\substack{i=1 \\i\neq k}}^K |\mathbf{h}_{k, n_{\text{f}}}^{\text{H}}\mathbf{w}_{i}|^2 + \sum_{n_0 = K+1}^{N_0^2} |\mathbf{h}_{k, n_{\text{f}}}^{\text{H}} \mathbf{w}_{n_0}|^2 + \zeta^2_{\text{c}}},
\end{equation}
where $\mathbf{h}_{k, n_{\text{f}}} = [\mathbf{h}^{\text{T}}_{1, k, n_{\text{f}}}, \ldots, \mathbf{h}^{\text{T}}_{N_{\text{b}}, k, n_{\text{f}}}]^{\text{T}}\in\mathbb{C}^{N_{\text{b}}N_0^2}$, and $\mathbf{w}_{k} = [\mathbf{w}^{\text{T}}_{1, k}, \ldots, \mathbf{w}^{\text{T}}_{N_{\text{b}}, k}]^{\text{T}}\in\mathbb{C}^{N_{\text{b}}N_0^2}$.
Finally, the spectral efficiency (SE) of the communication system is expressed as
\begin{equation}\label{eq-user-se}
\text{SE} = \sum_{k=1}^K\sum_{n_{\text{f}}=1}^{N_{\text{f}}}\log_2(1 + \gamma_{k, n_{\text{f}}}).
\end{equation}

\subsection{Sensing Model}

When the $n_{\text{b1}}$-th BS transmits $\mathbf{x}_{n_{\text{b1}}, n_{\text{f}}, n_{\text{s}}}$, the signal is scattered by UAVs and received by the $n_{\text{b2}}$-th BS. The received signal is given by
\begin{equation}\label{eq-sensing-signal-model-1}
\mathbf{r}^{\text{s}}_{n_{\text{b}1},n_{\text{b}2},n_{\text{f}},n_{\text{s}}}=\mathbf{F}_{n_{\text{b}2}}\mathbf{H}_{n_{\text{b}1},n_{\text{b}2},n_{\text{f}}}\mathbf{x}_{n_{\text{b1}}, n_{\text{f}}, n_{\text{s}}}+\tilde{\mathbf{z}}^{\text{s}}_{n_{\text{b}1},n_{\text{b}2},n_{\text{f}},n_{\text{s}}},
\end{equation}
where $\mathbf{H}_{n_{\text{b}1},n_{\text{b}2},n_{\text{f}}}\in \mathbb{C}^{N_0^2\times N_0^2}$ represents the channel from the $n_{\text{b}1}$-th BS, scattered by the targets in the ROI, and received at the $n_{\text{b2}}$-th BS.
The matrix $\mathbf{F}_{n_{\text{b}2}}$ is the combiner at the $n_{\text{b2}}$-th BS, designed to filter out direct signals and retain only scattered signals from the ROI.
By employing advanced beamforming methods \cite{li2023towards}, the sensing beampattern can be accurately oriented to cover the ROI with significantly suppressed sidelobes, making the interference signal power originating from targets outside the ROI and from the LOS path between BSs negligible \cite{he2024device}.
Additionally, the background interference scattered by static buildings can be measured in a calibration process and eliminated from the received signals during algorithm implementation \cite{li2024radio}.
Furthermore, given the sparsity of the low-altitude space, where rare scatterers may result in limited interference, we focus only on the signals passing through $\mathbf{H}_{n_{\text{b}1},n_{\text{b}2},n_{\text{f}}}$ in \eqref{eq-sensing-signal-model-1}, while other potential multipath components are treated as part of the additive noise $\tilde{\mathbf{z}}_{n_{\text{b}1},n_{\text{b}2},n_{\text{f}},n_{\text{s}}}^{\text{s}}$ \cite{liu2024cooperative,wang2024heterogeneous}.

The $(n_{\text{t}},n_{\text{r}})$-th element in $\mathbf{H}_{n_{\text{b}1},n_{\text{b}2},n_{\text{f}}}$, representing the channel between the $n_{\text{t}}$-th transmitting antenna of the $n_{\text{b1}}$-th BS and the $n_{\text{r}}$-th receiving antenna of the $n_{\text{b2}}$-th BS on the $n_{\text{f}}$-th subcarrier, is given by \cite{goldsmith2005wireless,huang2024fourier}:
\begin{multline}\label{eq-continuous-sensing-channel-model} 
h^{\text{s}}_{n_{\text{t}},n_{\text{r}},n_{\text{f}}} = \iiint \frac{\lambda_0 \sqrt{G_{\text{s}}}}{\sqrt{4\pi}} \times \frac{\breve{\sigma}(x, y, z)}{4 \pi d_{1}(x,y,z) d_{2}(x,y,z)} \\
\times e^{-j 2\pi\frac{d_{1}(x,y,z)+d_{2}(x,y,z)}{\lambda_{n_{\text{f}}}}} dxdydz, 
\end{multline}
where $G_{\text{s}}$ is the combined antenna gain at the transmitter and receiver.
$\breve{\sigma}(x, y, z)$ represents the continuous ROI image, where each value corresponds to the scattering coefficient at position $[x, y, z]^{\text{T}}$.
$d_1(x,y,z)$ and $d_2(x,y,z)$ are the distances from the scattering point to the $n_{\text{t}}$-th transmitting antenna and the $n_{\text{r}}$-th receiving antenna, respectively.
$\lambda_{n_{\text{f}}}$ is the wavelength of the $n_{\text{f}}$-th subcarrier.
$h^{\text{s}}_{n_{\text{t}},n_{\text{r}},n_{\text{f}}}$ includes the multipath channels scattered by all targets in the ROI.

Given the relatively slow velocity of UAVs, we assume that their positions and the channel $\mathbf{H}_{n_{\text{b}1},n_{\text{b}2},n_{\text{f}}}$ remain constant over $N_{\text{s}}=N_0^2$ ISAC symbol intervals \cite{cheng2024networked,liu2024cooperative}.\footnote{
For example, with a subcarrier spacing of 60 kHz, a UAV velocity of 20 m/s, and $N_0 = 5$, the required coherent channel time is 0.4 ms, corresponding to an 8 mm UAV displacement, which is extremely small compared to the voxel and ROI sizes.}
By stacking the $N_{\text{s}}$ received signals, we can obtain
\begin{equation}\label{eq-sensing-signal-model-2}
\mathbf{R}^{\text{s}}_{n_{\text{b}1},n_{\text{b}2},n_{\text{f}}}=\mathbf{F}_{n_{\text{b}2}}\mathbf{H}_{n_{\text{b}1},n_{\text{b}2},n_{\text{f}}}\mathbf{X}_{n_{\text{b1}}, n_{\text{f}}}+\tilde{\mathbf{Z}}^{\text{s}}_{n_{\text{b}1},n_{\text{b}2},n_{\text{f}}},
\end{equation}
where $\mathbf{R}^{\text{s}}_{n_{\text{b}1},n_{\text{b}2},n_{\text{f}}} = [\mathbf{r}^{\text{s}}_{n_{\text{b}1},n_{\text{b}2},n_{\text{f}},1}, \ldots, \mathbf{r}^{\text{s}}_{n_{\text{b}1},n_{\text{b}2},n_{\text{f}},N_{\text{s}}}]$, and $\mathbf{X}_{n_{\text{b}1},n_{\text{f}}}$ and $\tilde{\mathbf{Z}}^{\text{s}}_{n_{\text{b}1},n_{\text{b}2},n_{\text{f}}}$ are similarly defined.
Given the high degrees of design freedom in $\mathbf{F}_{n_{\text{b}2}}$ and $\mathbf{X}_{n_{\text{b1}}, n_{\text{f}}}$, they can be full-rank matrices.
Consequently, by transmitting $\mathbf{R}^{\text{s}}_{n_{\text{b}1},n_{\text{b}2},n_{\text{f}}}$ to the central processing unit (CPU), $\mathbf{H}_{n_{\text{b}1},n_{\text{b}2},n_{\text{f}}}$ can be estimated using the least squares (LS) method \cite{liu2024cooperative}, given as
\begin{equation}\label{eq-LS-channel-estimation}
\widehat{\mathbf{H}}_{n_{\text{b}1},n_{\text{b}2},n_{\text{f}}}=\mathbf{F}^{-1}_{n_{\text{b}2}}\mathbf{R}^{\text{s}}_{n_{\text{b}1},n_{\text{b}2},n_{\text{f}}}\mathbf{X}_{n_{\text{b}1},n_{\text{f}}}^{-1}.
\end{equation}

Assuming that each BS's transmitted sensing signals are received by all BSs, both monostatic and multi-static sensing modes can be realized. Compared to solely monostatic \cite{guan20213} or bistatic \cite{liu2024cooperative} sensing modes, this approach enables more comprehensive CSI measurements.
In the subsequent sections, the CSI measurement $\widehat{\mathbf{H}}_{n_{\text{b}1},n_{\text{b}2},n_{\text{f}}}$ is utilized for low-altitude image reconstruction, realizing flight activity surveillance.

\section{CS-Based Problem Formulation and Analysis}
\label{sec-problem-formulation}

In this section, we formulate the low-altitude surveillance problem using CS techniques, allowing for the simultaneous detection of multiple UAVs within a cooperative ISAC network. Furthermore, we evaluate the system’s sensing capabilities by deriving the PSF.
Finally, we analyze the off-grid errors, which are critical for understanding the limitations and potential improvements in imaging performance. 

\subsection{On-Grid Problem Formulation}
\label{sec-on-grid-problem-formulation}

\begin{figure}
    \centering
    \includegraphics[width=0.75\linewidth]{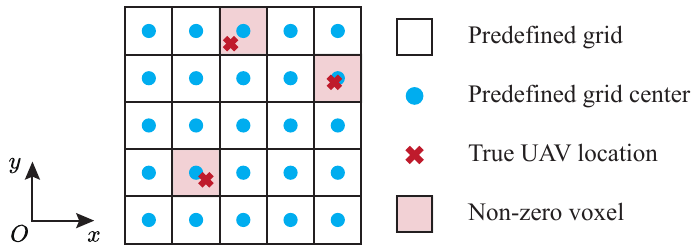}
    \captionsetup{font=footnotesize}
    \caption{2D illustration of low-altitude space discretization in the $xOy$ plane.}
    \label{fig-discretization}
\end{figure}

In \eqref{eq-continuous-sensing-channel-model}, the low-altitude image is represented as the continuous function $\breve{\sigma}(x, y, z)$. To enable aerial surveillance, we reconstruct the ROI image by discretizing it into $N_{\text{v}} = N_{\text{x}}\times N_{\text{y}}\times N_{\text{z}}$ voxels, each with a size of $d_{\text{x}}\times d_{\text{y}}\times d_{\text{z}}$.
Consequently, $\breve{\sigma}(x, y, z)$ is sampled into an $N_{\text{v}}$-dimensional vector $\boldsymbol{\sigma} = [\sigma_1, \ldots, \sigma_{N_{\text{v}}}]^{\text{T}}$, representing the unknown image to be estimated. A 2D slice of the low-altitude image in the $xOy$ plane is shown in Fig.~\ref{fig-discretization}.
We assume that $M$ UAVs, modeled as point targets \cite{zhao2024buptcmcc}, are randomly located within the ROI, occupying $M$ voxels, where $M\ll N_{\text{v}}$.
The scattering coefficient of the $n_{\text{v}}$-th voxel, denoted as $\sigma_{n_{\text{v}}}$, characterizes the UAV's scattering property if a UAV is present (pink voxels in Fig.~\ref{fig-discretization}). Otherwise, $\sigma_{n_{\text{v}}}=0$ (white voxels in Fig.~\ref{fig-discretization}).

Initially, we assume UAVs are exactly located at predefined voxel centers when $\sigma_{n_{\text{v}}}>0$, treating them as ``on-grid'' scatterers \cite{tong2021joint}.
This assumption facilitates modeling and performance analysis, providing insights for system configuration. However, as illustrated in Fig.~\ref{fig-discretization}, the true UAV location may deviate from voxel centers, introducing ``off-grid'' errors, which are analyzed in Sec. \ref{sec-off-grid-error-analysis} and addressed in Sec. \ref{sec-imaging-algorithm}.

\begin{figure}
\centering
\captionsetup{font=footnotesize}
\begin{subfigure}[b]{0.45\linewidth}
\centering
\includegraphics[width=\linewidth]{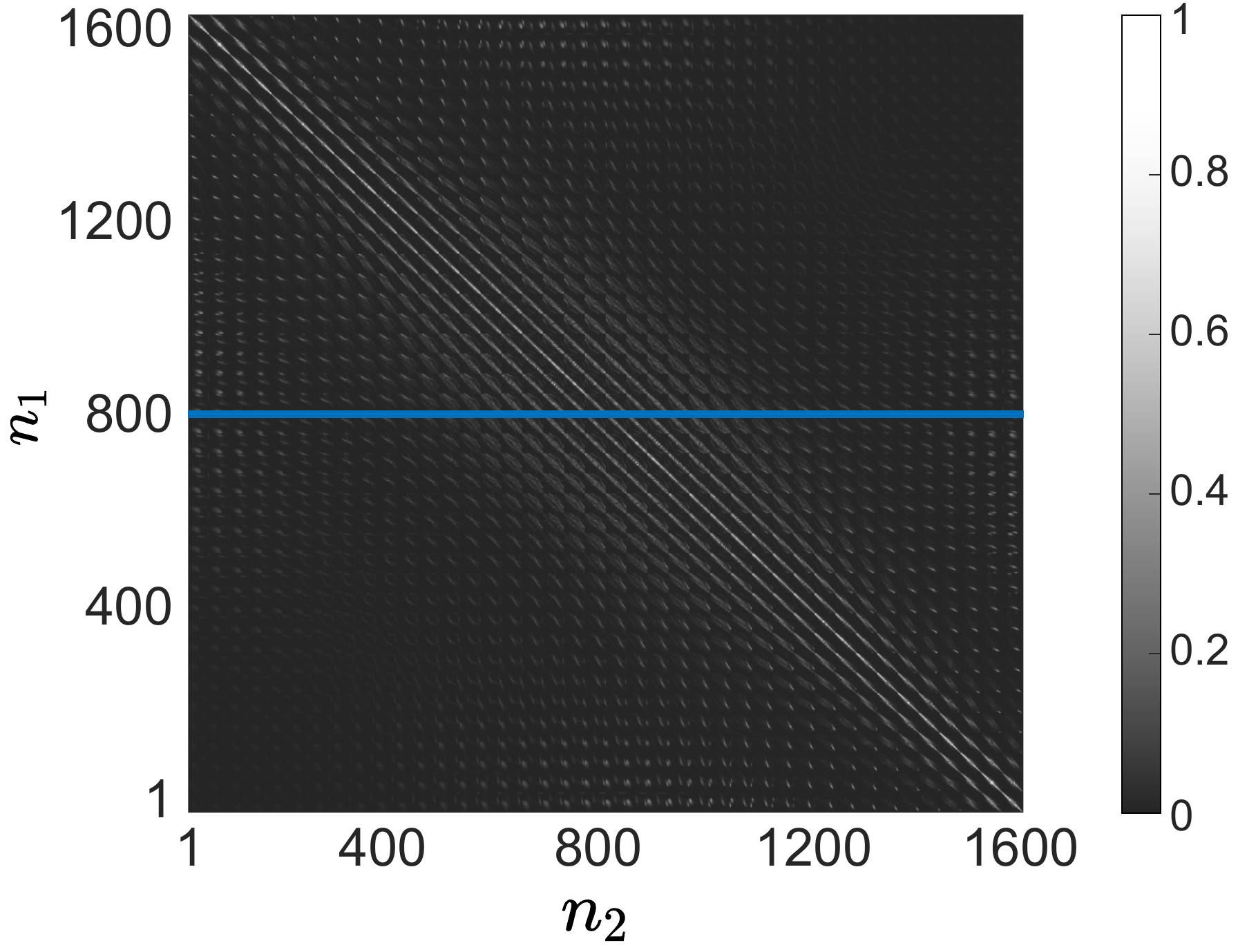}
\caption{}
\label{fig-psf-illustrate-1}
\end{subfigure}
\quad
\begin{subfigure}[b]{0.4\linewidth}
\centering
\includegraphics[width=\linewidth]{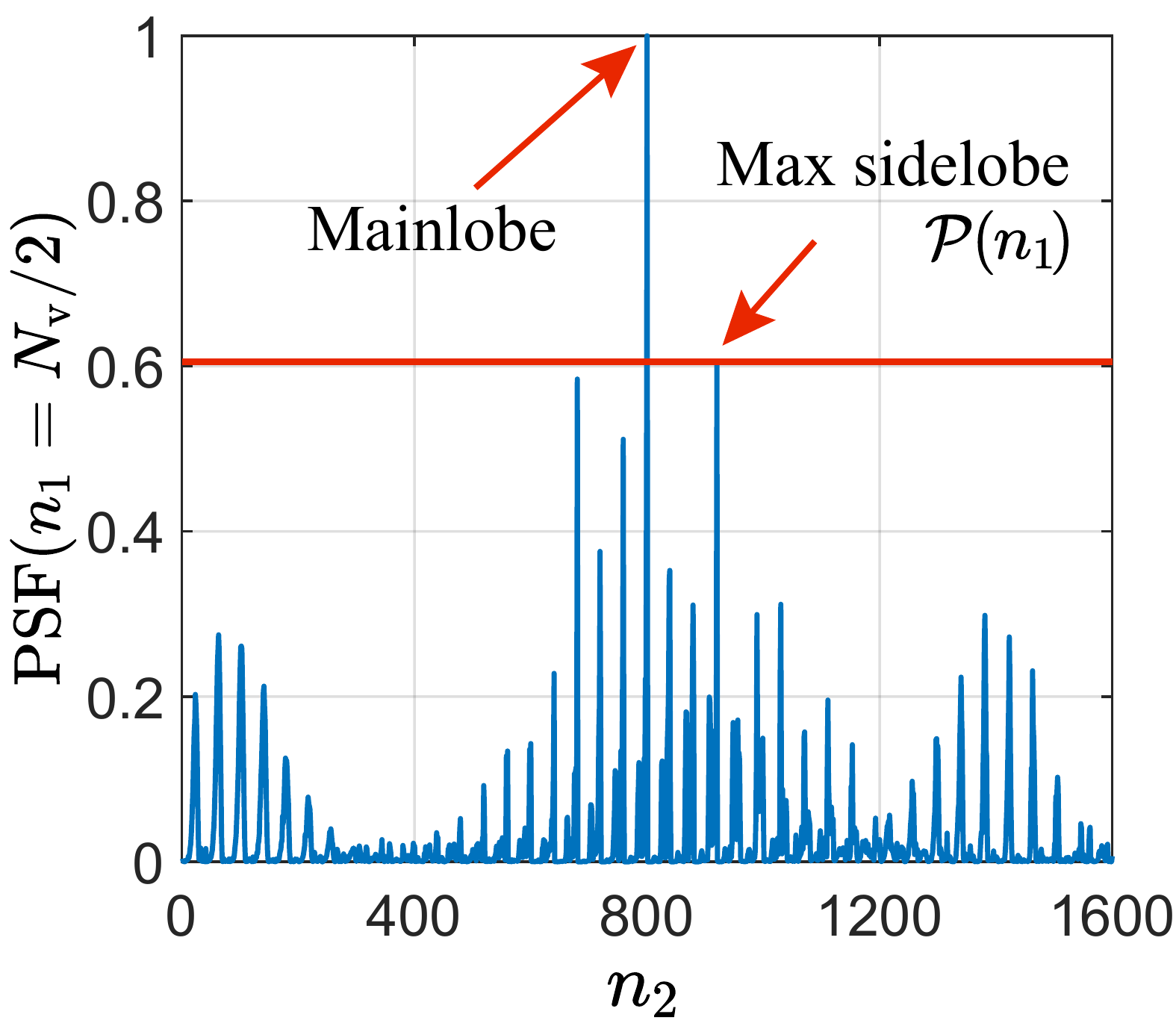}
\caption{}
\label{fig-psf-illustrate-2}
\end{subfigure}
\caption{Illustration of the 2D PSF (a) and its 1D slice (b), with (b) representing the blue line in (a).}
\label{fig-psf-illustrate}
\end{figure}

According to the cascaded channel model, the discrete form of \eqref{eq-continuous-sensing-channel-model} is given as \cite{tong2021joint,goldsmith2005wireless,huang2024fourier}
\begin{align}
h^{\text{s}}_{n_{\text{t}},n_{\text{r}},n_{\text{f}}}&=\sum_{n_{\text{v}}=1}^{N_{\text{v}}} \frac{\lambda_0 \sqrt{G_{\text{s}}}}{\sqrt{4\pi}} h_{n_{\text{t}},n_{\text{v}},n_{\text{f}}}\,\sigma_{n_{\text{v}}}\, h_{n_{\text{r}},n_{\text{v}},n_{\text{f}}} \notag \\
&=\frac{\lambda_0 \sqrt{G_{\text{s}}}}{\sqrt{4\pi}}\mathbf{h}_{n_{\text{t}},n_{\text{f}}}^{\text{H}}\text{diag}(\boldsymbol{\sigma})\mathbf{h}_{n_{\text{r}},n_{\text{f}}},
\label{eq-discretized-channel-model}
\end{align}
where $\mathbf{h}_{n_{\text{t}},n_{\text{f}}} \!\!=\!\! [h_{n_{\text{t}},1,n_{\text{f}}}, \ldots, h_{n_{\text{t}},N_{\text{v}},n_{\text{f}}}]^{\text{T}}$, with each element given by 
\begin{equation}\label{eq-free-space-channel}
h_{n_{\text{t}},n_{\text{v}},n_{\text{f}}}=\frac{e^{-j2\pi {d_{n_{\text{t}},n_{\text{v}}}}/{\lambda_{n_{\text{f}}}}}}{\sqrt{4\pi} d_{n_\text{t},n_{\text{v}}}},
\end{equation}
where $d_{n_{\text{t}},n_{\text{v}}}$ represents the distance from the $n_{\text{t}}$-th transmitting antenna to the $n_{\text{v}}$-th voxel.
Similar definitions apply for $\mathbf{h}_{n_{\text{r}},n_{\text{f}}}$ and $h_{n_{\text{r}},n_{\text{v}},n_{\text{f}}}$.
Consequently, the $(n_{\text{t}},n_{\text{r}})$-th element of $\widehat{\mathbf{H}}_{n_{\text{b}1},n_{\text{b}2},n_{\text{f}}}$ is given by 
\begin{equation}\label{eq-measurement-model-1}
\begin{aligned}
y_{n_{\text{t}},n_{\text{r}},n_{\text{f}}} = h^{\text{s}}_{n_{\text{t}},n_{\text{r}},n_{\text{f}}} + z_{n_{\text{t}},n_{\text{r}},n_{\text{f}}} = \mathbf{a}^{\text{H}}_{n_{\text{t}},n_{\text{r}},n_{\text{f}}}\boldsymbol{\sigma} + z_{n_{\text{t}},n_{\text{r}},n_{\text{f}}},
\end{aligned}
\end{equation}
where $\mathbf{a}^{\text{H}}_{n_{\text{t}},n_{\text{r}},n_{\text{f}}}=\frac{\lambda_0 \sqrt{G_{\text{s}}}}{\sqrt{4\pi}}\mathbf{h}_{n_{\text{t}},n_{\text{f}}}^{\text{H}}\text{diag}(\mathbf{h}_{n_{\text{r}},n_{\text{f}}})$, and $z_{n_{\text{t}},n_{\text{r}},n_{\text{f}}}$ represents additive noise from channel estimation.
By aggregating the measurements from all transmitting antennas, receiving antennas, and subcarriers, the CSI measurements related to the $n_{\text{b1}}$-th BS transmitter and the $n_{\text{b2}}$-th BS receiver is expressed as
\begin{equation}\label{eq-measurement-model-2}
\mathbf{y}_{n_{\text{b}1},n_{\text{b}2}}=\mathbf{A}_{n_{\text{b}1},n_{\text{b}2}}\boldsymbol{\sigma} + \mathbf{z}_{n_{\text{b}1},n_{\text{b}2}},
\end{equation}
where $\mathbf{A}_{n_{\text{b}1},n_{\text{b}2}}\in\mathbb{C}^{N_{\text{f}}N_0^4\times N_{\text{v}}}$, with its $(n_{\text{t}},n_{\text{r}},n_{\text{f}})$-th row given by $\mathbf{a}^{\text{H}}_{n_{\text{t}},n_{\text{r}},n_{\text{f}}}$.
The cellular network with $N_{\text{b}}$ BSs can effectively generate $N_{\text{b}}(N_{\text{b}}+1)/2$ independent measurement sets, each adhering to the model in \eqref{eq-measurement-model-2}. By stacking all measurements, the system equation is given by
\begin{equation}\label{eq-measurement-model-3}
\mathbf{y}=\mathbf{A}\boldsymbol{\sigma} + \mathbf{z},
\end{equation}
where $\mathbf{A}\in\mathbb{C}^{N_{\text{f}}N_0^4N_{\text{b}}(N_{\text{b}}+1)/2\times N_{\text{v}}}$ is the overall sensing matrix, and $\mathbf{z}$ is modeled as zero-mean AWGN with variance $\zeta^2_{\text{s}}$.

Our objective is to reconstruct the image $\boldsymbol{\sigma}$ from the measurement $\mathbf{y}$ using the sensing matrix $\mathbf{A}$.
Given that UAVs occupy only a small fraction of the total voxels ($M\ll N_{\text{v}}$), $\boldsymbol{\sigma}$ exhibits high sparsity. We leverage CS theory to formulate the low-altitude sensing problem, given as \cite{dai2009subspace,tong2021joint}:
\begin{equation}\label{eq-CS-problem}
\text{(P1)} \ \ \hat{\boldsymbol{\sigma}} = \argmin_{\boldsymbol{\sigma}} \|\boldsymbol{\sigma}\|_1, \ \ \text{s.t.} \ \|\mathbf{y}-\mathbf{A}\boldsymbol{\sigma}\|^2\le\varepsilon,
\end{equation}
where $\varepsilon$ is a small threshold ensuring reconstruction accuracy. Note that problem (P1) differs from traditional CS problems in two key aspects. First, the sensing matrix $\mathbf{A}$ may have a high condition number due to channel correlations among compactly arranged antennas. Second, its row count, determined by $N_0$, $N_{\text{b}}$, and $N_{\text{f}}$, may exceed its column count $N_{\text{v}}$, depending on system configurations. The abundance of measurements enhances low-altitude space sensing, mitigating uncertainties from $\mathbf{A}$'s large condition number and improving sparse vector recovery. 

\begin{figure*}
\centering
\captionsetup{font=footnotesize}
\begin{subfigure}[b]{0.28\linewidth}
\centering
\includegraphics[width=\linewidth]{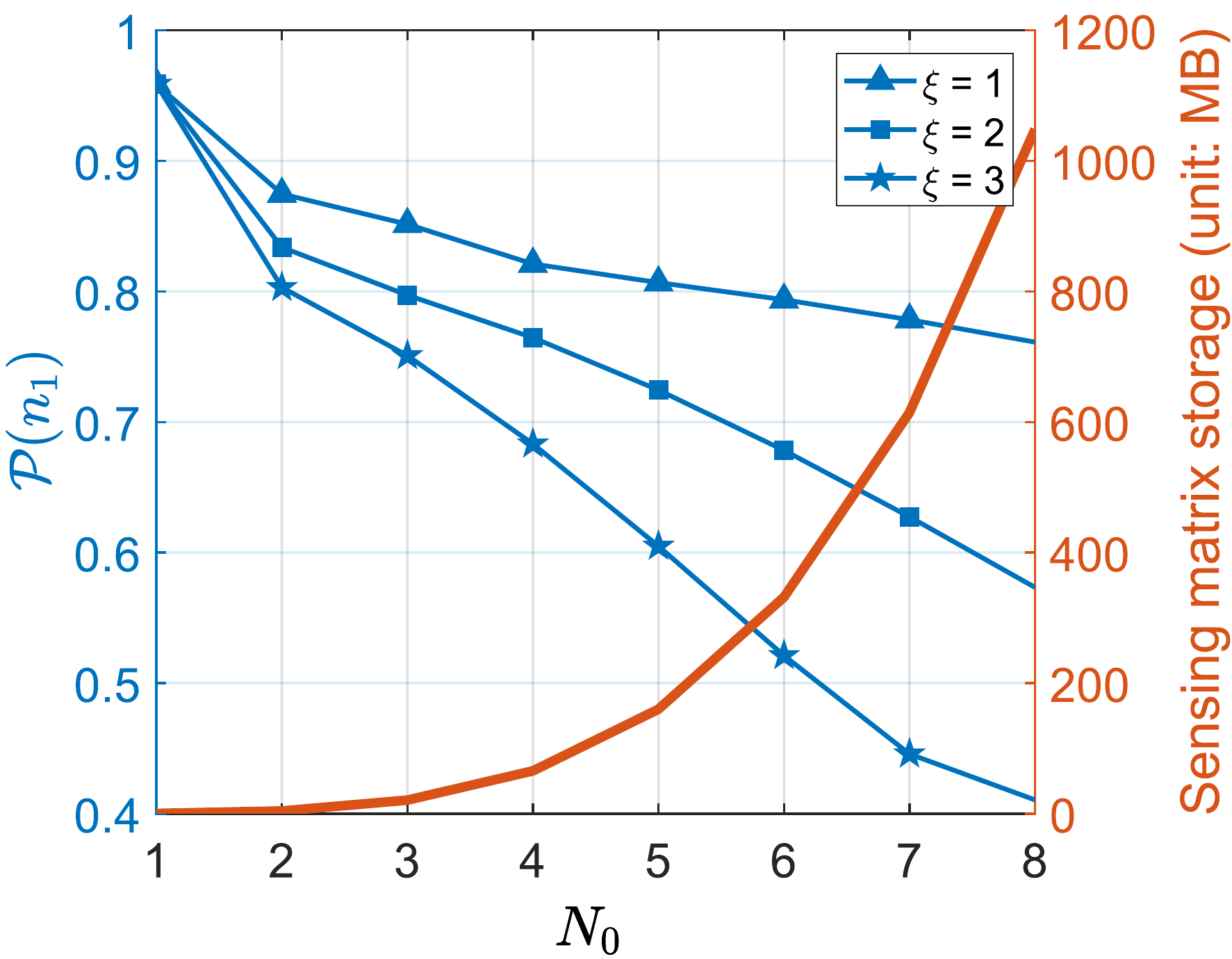}
\caption{}
\label{fig-psf-1}
\end{subfigure}
\quad
\begin{subfigure}[b]{0.28\linewidth}
\centering
\includegraphics[width=\linewidth]{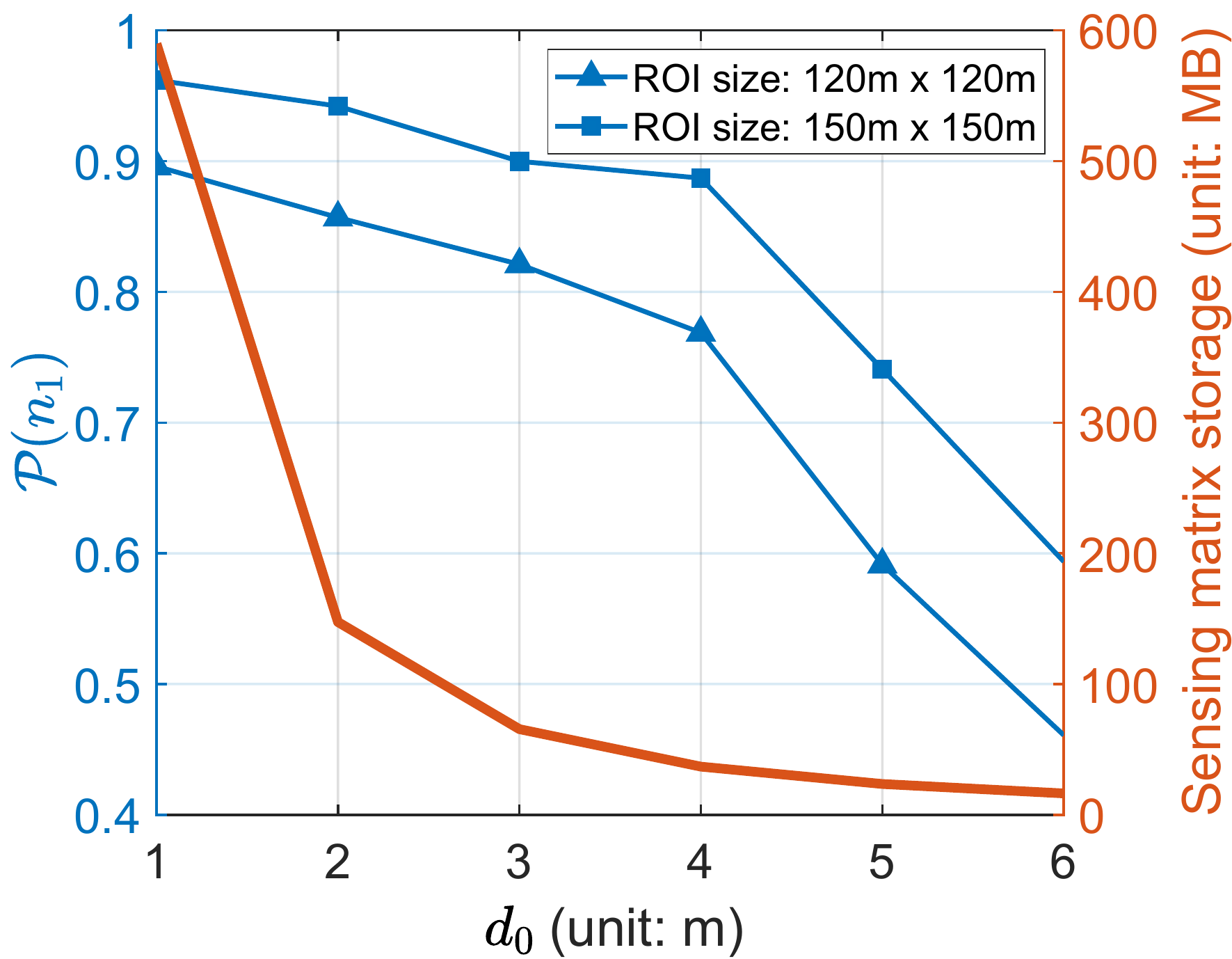}
\caption{}
\label{fig-psf-2}
\end{subfigure}
\quad
\begin{subfigure}[b]{0.28\linewidth}
\centering
\includegraphics[width=\linewidth]{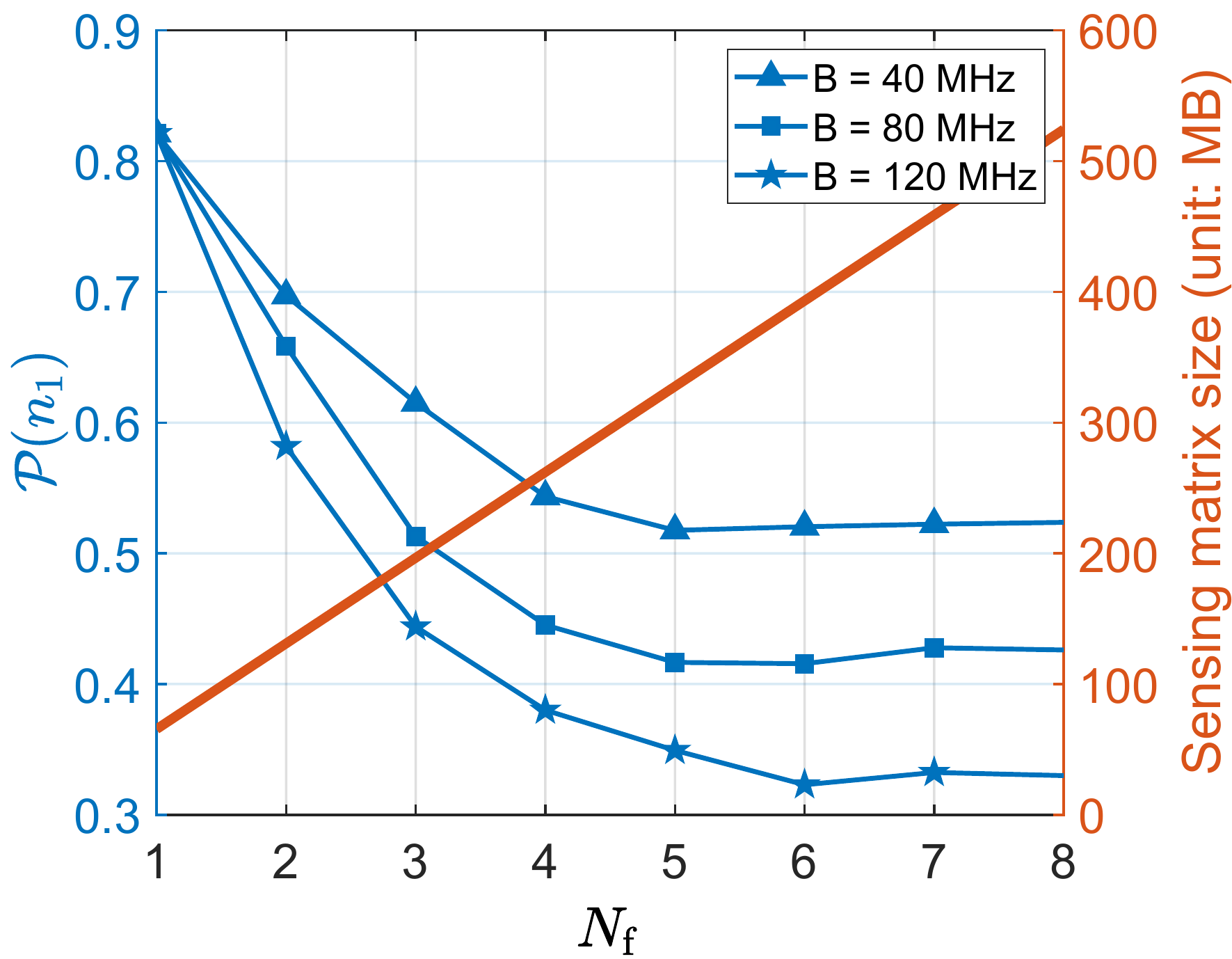}
\caption{}
\label{fig-psf-3}
\end{subfigure}
\caption{Maximum PSF sidelobe $\mathcal{P}(n_1)$ under varying system parameters: (a) Antenna number $N_0$ and antenna spacing scale $\xi$; (b) Voxel size $d_0$ and ROI size; (c) Subcarrier number $N_{\text{f}}$ and bandwidth $B$.}
\label{fig-psf}
\end{figure*}
 
\begin{remark}
The proposed problem formulation for low-altitude surveillance offers several advantages:
\begin{itemize}
    \item \textbf{Non-Cooperative Target Sensing}: UAV detection is achieved without any target cooperation.

    \item \textbf{Direct CSI Utilization}: CSI is directly used for image formation, eliminating the need for delay and angular parameter estimation as required in localization-based methods. This mitigates error propagation issues \cite{shih2024machine}.

    \item \textbf{Efficient Data Fusion}: Measurements from multiple BSs are directly stacked and fused in their raw form, avoiding the need for time-consuming beam scanning and complex decision-level image fusion \cite{ma2024networked,tang2024cooperative}.

    \item \textbf{Scalability}: Multiple targets can be detected simultaneously, with computational complexity independent of the number of targets.

    \item \textbf{Comprehensive Target Characterization}: Both target existence and scattering coefficients are estimated, providing valuable information for low-altitude surveillance. 
\end{itemize}
\end{remark} 

In this study, precise UAV localization is not the primary objective, since meter-to-ten-meter level accuracy is adequate for intrusion detection and trajectory planning \cite{ChinaMobile2024WhitePaper,3gpp2023study}. Instead, we focus on detecting non-zero voxels and estimating their scattering coefficients. The localization accuracy depends on voxel grid resolution, which should be chosen based on sensing capabilities and application needs, as discussed next.

\subsection{Sensing Ability Analysis with PSF}
\label{sec-psf}

The PSF is a target-independent metric used to evaluate the imaging capability of a given system and is widely employed in radar imaging systems with Fourier transform operations \cite{huang2024fourier}.
In \cite{patel2010compressed}, the PSF was adopted for CS-based imaging ability analysis by treating the columns of the sensing matrix as ``steering vectors''. Accordingly, the PSF is defined as 
\begin{equation}\label{eq-psf}
\operatorname{PSF}\left(n_{1}, n_{2}\right)=\frac{\left|\left\langle\mathbf{A}\left(:, n_{1}\right), \mathbf{A}\left(:, n_{2}\right)\right\rangle\right|}{\left\|\mathbf{A}\left(:, n_{1}\right)\right\|_{2}\left\|\mathbf{A}\left(:, n_{2}\right)\right\|_{2}},
\end{equation}
where $n_{1}, n_{2} = 1, 2, \ldots, N_{\text{v}}$, and $\mathbf{A}\left(:, n_{1}\right)$ represents the $n_{1}$-th column of $\mathbf{A}$.
The PSF quantifies the mutual coherence between steering vectors, where lower coherence indicates improved imaging capabilities. Thus, analyzing the PSF can provide guidelines for system configurations to achieve high sensing performance.

Using the simulation parameters described in Sec. \ref{sec-simu-on-1}, Fig.~\ref{fig-psf-illustrate-1} presents the PSF with $N_{\text{v}} = 1600$, revealing strong correlations among the steering vectors of adjacent voxels. A one-dimensional PSF slice with $n_1 = 800$ is shown in Fig.~\ref{fig-psf-illustrate-2}. According to CS theory, orthogonal columns in the sensing matrix enable sparse vector recovery with minimal uncertainty. However, high channel correlations among adjacent antennas and subcarriers result in elevated PSF sidelobes, as depicted in Fig.~\ref{fig-psf-illustrate}, thereby making voxel discrimination challenging.
To achieve high-resolution and high-accuracy imaging, \emph{small voxel sizes} and \emph{large measurement sets} are generally preferred. However, these configurations may lead to increased PSF sidelobes and expanded sensing matrix dimensions, which in turn reduce imaging accuracy and impose substantial computational burdens.

We next analyze the effects of different system parameters on the maximum PSF sidelobe, defined as $\mathcal{P}(n_1) = \max(\operatorname{PSF}(n_1, n_2))$, where $n_1 \neq n_2$. A lower $\mathcal{P}(n_1)$ typically indicates improved imaging performance. Key observations from the simulation results in Fig.~\ref{fig-psf} are summarized as follows:

\begin{enumerate}
\item \textbf{Antenna Configuration (Fig.~\ref{fig-psf-1})}:  Increasing $N_0$ reduces $\mathcal{P}(n_1)$ by aggregating more CSI measurements for image reconstruction. However, the sensing matrix size increases exponentially, surpassing 1 GB storage at $N_0 = 8$, which results in high computational complexity. Enlarging the antenna spacing to $(\lambda_0/2) \cdot \xi$ significantly lowers $\mathcal{P}(n_1)$, indicating that sparse arrays enhance spatial resolution and improve low-altitude imaging performance.

\item \textbf{Voxel Size and ROI Coverage (Fig.~\ref{fig-psf-2})}: 
When $d_0 = d_{\text{x}} = d_{\text{y}} = d_{\text{z}}$, larger voxel sizes reduce PSF sidelobes by increasing the spacing between voxels and decreasing channel correlation. However, this comes at the cost of reduced image resolution, although the sensing matrix becomes smaller. Therefore, a trade-off exists between imaging resolution and reconstruction accuracy. 

\item \textbf{Subcarrier and Bandwidth Configuration (Fig.~\ref{fig-psf-3})}: Increasing the bandwidth and number of subcarriers improves PSF performance by enriching the available information from the ROI, thereby lowering $\mathcal{P}(n_1)$. However, if the bandwidth remains fixed while $N_{\text{f}}$ increases, $\mathcal{P}(n_1)$ eventually saturates due to decreased subcarrier spacing and increased channel correlation. As a result, further increases in $N_{\text{f}}$ do not necessarily enhance imaging accuracy but substantially raise computational costs. 
\end{enumerate}
These findings provide practical guidelines for configuring ISAC networks, facilitating a balance among imaging resolution, reconstruction accuracy, and computational efficiency. Additional analysis involving other system parameters can be conducted using the PSF formulation given in \eqref{eq-psf}. 

\subsection{Off-Grid Error Analysis and Problem Reformulation}
\label{sec-off-grid-error-analysis}

Sec. \ref{sec-on-grid-problem-formulation} formulated the on-grid imaging problem, where the sensing matrix $\mathbf{A}$ was constructed using predefined voxel center positions. However, only on-grid targets (blue points in Fig.~\ref{fig-discretization}) can be accurately detected, as their steering vectors are included in $\mathbf{A}$. As illustrated in Fig.~\ref{fig-discretization}, UAVs rarely align perfectly with the predefined grid, introducing off-grid errors where the steering vectors in $\mathbf{A}$ fail to match the true UAV positions. 
Specifically, $\sigma_{n_{\text{v}}}>0$ only states that the $n_{\text{v}}$-th voxel include a UAV but it may not locate at the voxel center.
Consequently, the equality in \eqref{eq-measurement-model-3} no longer holds.

\begin{figure}
    \centering
    \includegraphics[width=0.6\linewidth]{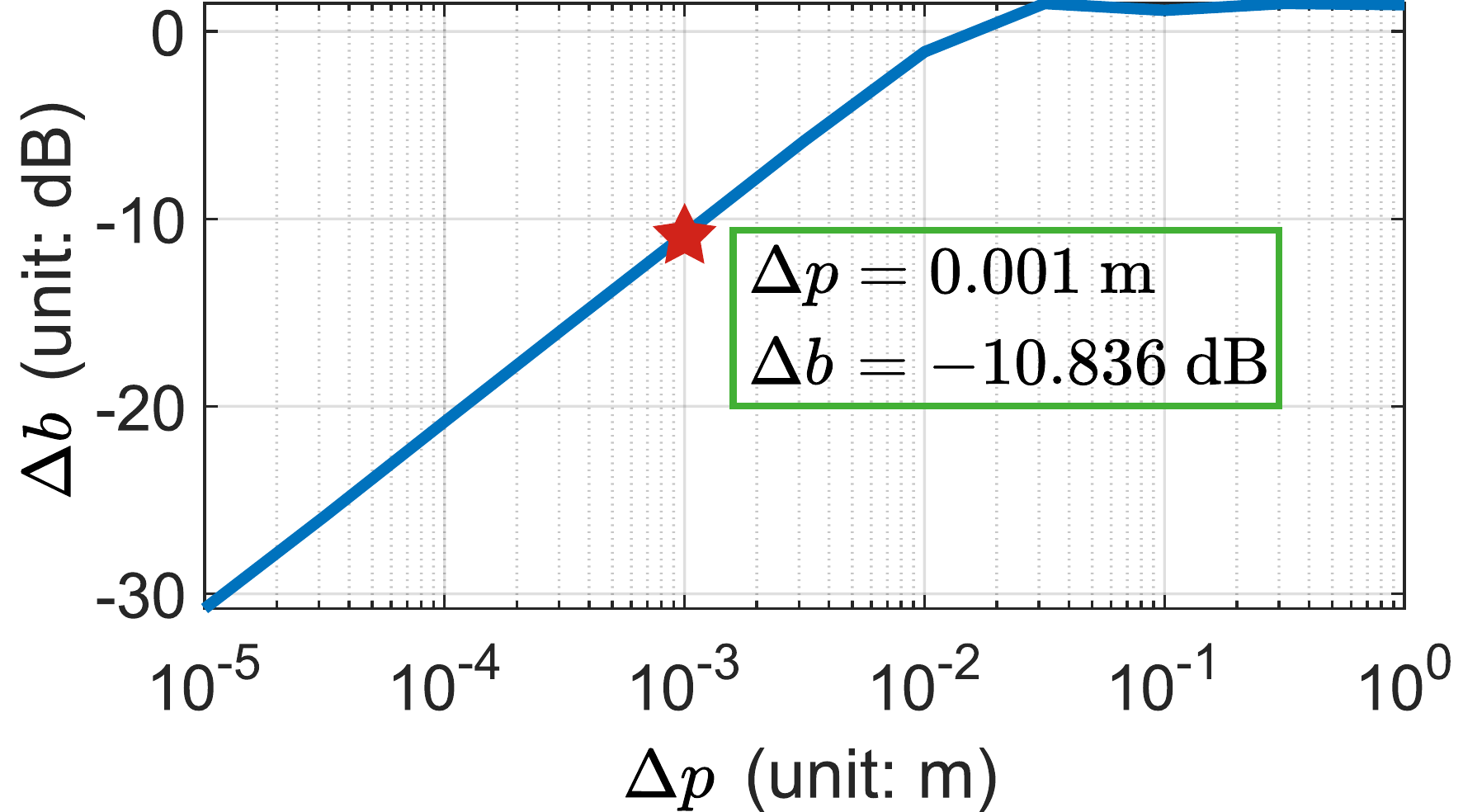}
    \captionsetup{font=footnotesize}
    \caption{Relationship between off-grid error $\Delta b$ and location discrepancy $\Delta p$.}
    \label{fig-off-grid-error}
    \vspace{-0.05cm}
\end{figure}

A direct approach to reduce off-grid errors is increasing grid density. Let $\mathbf{p}$ and $\mathbf{p}^*$ denote the true UAV location and the nearest grid center, respectively. Using the formulation in Sec. \ref{sec-on-grid-problem-formulation} and the simulation parameters in Sec. \ref{sec-simu-on-1}, we compute the corresponding steering vectors $\mathbf{b}$ and $\mathbf{b}^*$, respectively. The off-grid error can then be quantified by
\begin{equation}
\Delta b = {\|\mathbf{b} - \mathbf{b}^*\|_2}/{\|\mathbf{b}\|_2}.
\end{equation}
Fig.~\ref{fig-off-grid-error} depicts the relationship between $\Delta b$ and the location discrepancy $\Delta p = \|\mathbf{p} - \mathbf{p}^*\|_2$.
The results show that reducing $\Delta b$ below $-10\ \text{dB}$ requires $\Delta p<1\ \text{mm}$. Achieving such dense grids results in 1) large sensing matrices and high computational costs, and 2) amplified PSF sidelobes (Fig.~\ref{fig-psf-2}), reducing imaging accuracy. Thus, increasing grid density alone is impractical for mitigating off-grid errors.

The above analysis highlights that off-grid errors introduce significant challenges for low-altitude imaging. Specifically, the constraint in problem (P1) becomes ineffective due to the invalidity of \eqref{eq-measurement-model-3}. Moreover, existing off-grid formulations based on Taylor expansion or ANM \cite{yang2012off,wei2022off,li2024atomic,gao2024robust} are unsuitable for low-altitude imaging due to their complex mathematical structures and high computational costs.
To address the off-grid issues, we generalize the imaging problem as follows:
\begin{equation}
\text{(P2)} \ \ \hat{\boldsymbol{\sigma}} = \argmin_{\boldsymbol{\theta}} \|\hat{\boldsymbol{\sigma}}-\boldsymbol{\sigma}\|_2, \ \ \text{s.t.} \ \hat{\boldsymbol{\sigma}}=f_{\boldsymbol{\theta}}(\mathbf{y}, \mathbf{A}),
\end{equation}
where $\hat{\boldsymbol{\sigma}}$ is the estimated sparse image, and $f_{\boldsymbol{\theta}}(\mathbf{y}, \mathbf{A})$ is an imaging function parameterized by $\boldsymbol{\theta}$.
The off-grid issues have been released in the objective function of (P2), since it only cares which voxel includes a UAV rather than the UAV's exact location.
Despite these modifications, the on-grid imaging models introduced in Sec. \ref{sec-on-grid-problem-formulation} remain essential for PSF-based analysis shown in Sec. \ref{sec-psf} and algorithm development discussed in the next section.

\section{Imaging Algorithms}
\label{sec-imaging-algorithm}

The previous section formulated on-grid and off-grid imaging problems for low-altitude surveillance. In this section, we first introduce a traditional CS-based on-grid imaging algorithm to solve (P1). Then, to address (P2), we propose a learned physics-embedded off-grid imager, incorporating novel loss functions based on the OHEM scheme.

\subsection{CS-based On-Grid Imaging Algorithm}
\label{sec-algo-on-grid}

Various algorithms can be employed to solve (P1) \cite{zou2022concise}. Considering the tradeoff between estimation accuracy, computational complexity, and prior knowledge requirements, we adopt the subspace pursuit (SP) algorithm \cite{dai2009subspace}, which incorporates iterative refinement to improve reconstruction accuracy.

We define the residual signal computation as
\begin{equation}
\mathbf{y}_{\text{res}}(\mathcal{S}) = \mathbf{y} - \mathbf{A}_{\mathcal{S}} f_{\text{LS}}(\mathbf{y}, \mathbf{A}_{\mathcal{S}}),
\end{equation}
where $\mathcal{S}$ represents the support set of the sparse signal, and $\mathbf{A}_{\mathcal{S}}$ is the corresponding sub-matrix of $\mathbf{A}$.
The function $f_{\text{LS}}(\mathbf{y}, \mathbf{A}_{\mathcal{S}})$ computes the LS estimate of the non-zero values in $\boldsymbol{\sigma}$.
Additionally, the function $f_{{\text{sel}},M}(\mathbf{y}, \mathbf{A})$ selects the indices corresponding to the largest $M$ absolute values of $\mathbf{A}^{\text{H}}\mathbf{y}$, effectively selecting the potential signal support.

The SP algorithm initializes by generating an initial support $\mathcal{S}_0$ using $f_{{\text{sel}},M}(\mathbf{y}, \mathbf{A})$ and computing the corresponding residual $\mathbf{y}_{\text{res}}(\mathcal{S}_0)$. In the $i$-th iteration, the following steps are performed: 
\begin{enumerate}
    \item \textbf{Expand the support}: Augment $\mathcal{S}_{i-1}$ by adding the indices obtained from $f_{{\text{sel}},M}(\mathbf{y}_{\text{res}}(\mathcal{S}_{i-1}), \mathbf{A})$, forming $\tilde{\mathcal{S}}_i$.
    \item \textbf{Update the support}: Refine the support as $\mathcal{S}_i=f_{{\text{sel}},M}(\mathbf{y}, \mathbf{A}_{\tilde{\mathcal{S}}_i})$.
    \item \textbf{Update the residual}: Compute the new residual $\mathbf{y}_{\text{res}}(\mathcal{S}_i)$.
\end{enumerate}
The process repeats until the residual falls below a threshold $\varepsilon$ or the support set stabilizes. Finally, the estimated non-zero values in $\boldsymbol{\sigma}$ are computed as $f_{\text{LS}}(\mathbf{y}, \mathbf{A}_{\mathcal{S}_{{\text{final}}}})$, where $\mathcal{S}_{{\text{final}}}$ denotes the final support set. Since the exact number of UAVs is unknown, the algorithm uses a prior-based sparsity value $M^\circ$.
The SP algorithm is summarized in Algorithm \ref{ag1}, with a computational complexity of $O(N_{\text{i}}(N_1N_{\text{v}} + N_1M^{\circ 2}))$, primarily due to the least squares (LS) estimation and matrix-vector multiplications. Here, $N_{\text{i}}$ denotes the number of iterations, and $N_1 = N_{\text{f}}N_0^4N_{\text{b}}(N_{\text{b}}+1)/2$ represents the total number of measurements. The value $M^{\circ}$ serves as the prior-based sparsity input to the SP algorithm.

\begin{algorithm}[t]
\caption{The SP algorithm \cite{dai2009subspace}.}
\label{ag1}
\begin{algorithmic}[1]
        \STATE $\mathbf{input:}$ $\mathbf{A}$, $\mathbf{y}$, and prior-based sparsity $M^\circ$.

        \STATE $\mathbf{initialize:}$ Calculate initial support $\mathcal{S}_0 = f_{{\text{sel}},M^\circ}(\mathbf{y}, \mathbf{A})$, derive the residual $\mathbf{y}_{\text{res}}(\mathcal{S}_0)$, and set $i=0$.

        \STATE $\mathbf{while} \ \|\mathbf{y}_{\text{res}}(\mathcal{S}_i)\|_2>\varepsilon \ \text{or}\ \mathcal{S}_{i-1} \ne \mathcal{S}_{i} \ (i \ge 1) \ \mathbf{do}$

        \STATE \hspace{0.5cm} $i=i+1$.

        \STATE \hspace{0.5cm} Derive $\tilde{\mathcal{S}}_i=\cup (\mathcal{S}_{i-1}, f_{{\text{sel}},M^\circ}(\mathbf{y}_{\text{res}}(\mathcal{S}_{i-1}), \mathbf{A}))$.

        \STATE \hspace{0.5cm} Renew the support as $\mathcal{S}_i=f_{{\text{sel}},M^\circ}(\mathbf{y}, \mathbf{A}_{\tilde{\mathcal{S}}_i})$.

        \STATE \hspace{0.5cm} Update the residual as $\mathbf{y}_{\text{res}}(\mathcal{S}_i)$.

        \STATE $\mathbf{end\ while}$

        \STATE $\mathbf{output:}$ the estimated ROI image $\hat{\boldsymbol{\sigma}}$.

\end{algorithmic}
\end{algorithm}

\subsection{Physics-Embedded Learning Under Off-Grid Conditions}
\label{sec-algo-off-grid}

\begin{figure}
    \centering
    \includegraphics[width=0.9\linewidth]{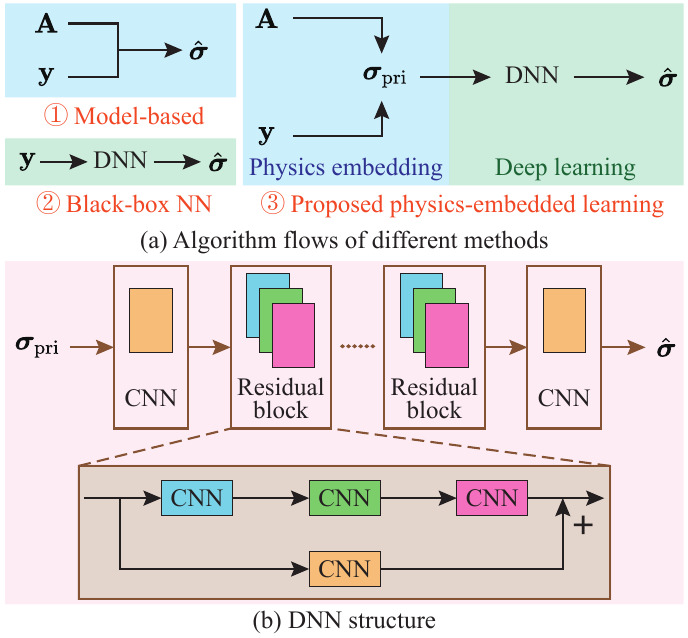}
    \captionsetup{font=footnotesize}
    \caption{Algorithm flow and DNN structure illustrations.}
    \label{fig-algorithm-flow}
\end{figure}

Traditional CS-based algorithms have demonstrated high performance in on-grid imaging tasks \cite{tong2021joint}. However, their effectiveness degrades under off-grid conditions due to modeling errors, as discussed in Sec. \ref{sec-off-grid-error-analysis}. To address (P2), we leverage DNNs while recognizing the limitations of Taylor expansion and ANM-based approaches for low-altitude imaging \cite{yang2012off,wei2022off,li2024atomic,gao2024robust}.
Since black-box DNNs often overlook physical constraints \cite{huang2023off,zhang2024two}, we adopt the physics-embedded learning framework \cite{guo2023physics,wu2019deep,chen2020review}, integrating model-based priors with data-driven learning to enhance imaging accuracy.
As shown in Fig.~\ref{fig-algorithm-flow}, the proposed approach comprises two stages: 

\subsubsection{Step 1---Primary Model-Based Processing}

Although the on-grid model does not perfectly represent the off-grid scenario, an initial estimate can be obtained by\footnote{
The primary result may also be obtained by using the pseudo inverse of $\mathbf{A}$.
However, the computational complexity of \eqref{eq-primary} is $O(N_{\text{v}}N_1)$ but $O(N_{\text{v}}^2(N_{\text{v}} + N_1))$ for pseudo-inverse calculations.
Furthermore, the high condition number of $\mathbf{A}$ may lead to results significantly distorted by the additive noise when using the pseudo inverse.}
\begin{equation}\label{eq-primary}
\boldsymbol{\sigma}_{\text{pri}} = \mathbf{A}^{\text{H}}\mathbf{y},
\end{equation}
where the matrix $\mathbf{A}$ can be calculated based on known BS antenna locations and predefined ROI grid positions.
This step is commonly used in CS-based algorithms \cite{dai2009subspace,zou2022concise,tong2021joint} for identifying sparse signal components. Unlike the SP algorithm, which applies thresholding and may discard useful information, this step directly projects the measurement data into the image domain without enforcing sparsity constraints.
While \eqref{eq-primary} provides an initial image, it lacks spatial accuracy due to off-grid errors in $\mathbf{A}$.

\subsubsection{Step 2---DNN Refinement}
To refine the initial estimate $\boldsymbol{\sigma}_{\text{pri}}$, we employ a DNN-based imaging model trained to reconstruct low-altitude images. As illustrated in Fig.~\ref{fig-algorithm-flow}, the DNN takes $\boldsymbol{\sigma}_{\text{pri}}$ as input and is trained with the ground truth $\boldsymbol{\sigma}$ as labels. The network learns to extract underlying image information from $\boldsymbol{\sigma}_{\text{pri}}$ and outputs a refined estimate 
$\hat{\boldsymbol{\sigma}}$, accurately detecting UAV locations in 3D space.
The DNN architecture integrates convolutional layers with residual connections, improving training convergence and mitigating issues like gradient vanishing and gradient explosion \cite{wu2019deep}.
Implementation of the DNN requires an offline training procedure using collected or simulated datasets, followed by an online inference step using optimized DNN parameters.
The computational complexity of the DNN inference is $O\left(\sum_{n_{\text{c}}=1}^{N_{\text{c}}} N_{\text{v}}N_{\text{k}, n_{\text{c}}}C_{\text{in}, n_{\text{c}}}C_{\text{out}, n_{\text{c}}}\right)$ \cite{nakata2021adaptive}, where $N_{\text{c}}$ is the number of CNN layers, $N_{\text{k}, n_{\text{c}}}$ is the number of variables in the convolutional kernel of the $n_{\text{c}}$-th layer, and $C_{\text{in}, n_{\text{c}}}$ and $C_{\text{out}, n_{\text{c}}}$ are the input and output channel counts, respectively. Although the computational complexity may be higher than that of Algorithm \ref{ag1}, the efficient computation capabilities of GPUs can enable near real-time inference after training.

\begin{remark}
Training data for real-world deployments can be collected using cooperative UAVs equipped with onboard localization devices. The CSI measurements can be acquired through signal transmission and reception by the ISAC network, while the ground-truth low-altitude image is generated based on the cooperative UAV locations.
The dataset can be reused across different scenarios, provided that the ISAC network topology and the relative positions of the ROI and BSs remain unchanged.
Additionally, synthetic datasets can be generated via computer simulations based on small real-world datasets, thereby reducing the need for extensive field data collection \cite{ohta2024real2sim2real}.
\end{remark}

\subsection{Loss Function Design based on OHEM}

To refine $\boldsymbol{\sigma}_{\text{pri}}$ towards the ground truth $\boldsymbol{\sigma}$, previous studies have employed the MSE loss function \cite{huang2023off,zhang2024two,wu2019deep}.
However, in low-altitude imaging, this approach faces significant challenges due to the extreme sparsity of $\boldsymbol{\sigma}$, where nearly all voxel values are zero, with only a few non-zero points. Consequently, the DNN may converge to an all-zero output, achieving relatively low MSE values but failing to detect UAVs. Furthermore, low-altitude images consist of isolated points rather than structured features, making it difficult for CNNs to extract meaningful spatial information. This increases the challenge of detecting UAVs and requires careful DNN training using specially designed loss functions.

To improve DRs and training efficiency, we adopt the simple and intuitive OHEM scheme \cite{shrivastava2016training} to design effective loss functions.
Originally developed to mitigate class imbalance in dataset, OHEM can help the DNN to focus on hard-to-detect targets, which are underrepresented in training samples.
Following the OHEM principle, voxels in each training image are categorized as: 
\begin{itemize}
    \item {\bf Positive Samples (Hard Samples):} Non-zero voxels representing UAVs (pink in Fig.~\ref{fig-discretization}), which form a small fraction of the total voxels.
    
    \item {\bf Negative Samples (Easy Samples):} Zero voxels representing empty space (white in Fig.~\ref{fig-discretization}), which dominate the voxel distribution.
\end{itemize}
For each predicted image, we define the loss contributions as follows:
\begin{itemize}
    \item {\bf Positive Sample Loss:} $L_{\text{pos}} = \sum_{\imath=1}^M l_{\text{pos}, \imath}$, where $l_{\text{pos}, \imath}$ represents the MSE loss of the $\imath$-th positive voxel. 

    \item {\bf Negative Sample Loss:} $L_{\text{neg}} = \sum_{\jmath=1}^{\eta M} l_{\text{neg}, \jmath}$, where $l_{\text{neg}, \jmath}$ is the MSE loss of the $\jmath$-th sorted negative voxel. Only the largest $\eta M$ MSE values of negative samples, determined by the hyper-parameter $\eta$, are selected. 
\end{itemize}

We introduce two OHEM-based loss functions:
\begin{equation}\label{eq-loss-ohem}
L_{\text{ohem}1} = \frac{L_{\text{pos}}+L_{\text{neg}}}{N_{\text{pos}}+N_{\text{neg}}},\quad L_{\text{ohem}2} = \frac{L_{\text{pos}}}{N_{\text{pos}}} + \frac{L_{\text{neg}}}{N_{\text{neg}}},
\end{equation}
where $N_{\text{pos}}=M$ and $N_{\text{neg}}=\eta M$ denote the numbers of positive and selected negative samples.
As a result, the model parameter update $\Delta \boldsymbol{\theta}$ in each training batch is mainly influenced by the limited positive samples and the selected negative samples with large MSE values. This satisfies $\Delta \boldsymbol{\theta} \propto -\nabla_{\boldsymbol{\theta}} L_{\text{ohem}\star}$, where $\star\in\{1,2\}$ and $\nabla_{\boldsymbol{\theta}}$ denotes the gradient with respect to $\boldsymbol{\theta}$.
Well-predicted negative samples are excluded from consideration. This prevents $\Delta \boldsymbol{\theta}$ from being dominated by a potentially large number of easy negatives and acts similarly to a thresholding operation.
By adjusting $\eta$, the value of $N_{\text{neg}}$ changes, which modifies the threshold that determines which negative samples are selected or excluded in $L_{\text{ohem}\star}$ during training.
Therefore, $\eta$ affects the optimization of $\boldsymbol{\theta}$ by balancing the influences of positive and negative samples. This results in different numbers of non-zero voxels in the predicted output.

However, the two loss functions in \eqref{eq-loss-ohem} behave differently as $\eta$ varies, requiring careful tuning.
For $L_{\text{ohem}1}$, MSE contributions from positive and selected negative samples ($l_{\text{pos}, \imath}$ and $l_{\text{neg}, \jmath}$) receive equal weight in backpropagation.
Increasing $N_{\text{neg}}$ amplifies the effect of $L_{\text{neg}}$, causing the network to favor zero voxel predictions. If all negative samples are selected, $L_{\text{ohem}1}$ converges to traditional MSE loss.
In contrast, $L_{\text{ohem}2}$ separately normalizes $L_{\text{pos}}$ and $L_{\text{neg}}$, giving both normalized terms equal weight. However, as $N_{\text{neg}}$ increases, the contribution of each negative sample diminishes, reducing the DNN’s bias toward zero predictions. This property makes $L_{\text{ohem}2}$ more effective for achieving higher DRs.

Given the sparse nature of low-altitude images, we can add a sparse regularization term to derive the final loss function:
\begin{equation}\label{eq-loss-final}
L = L_{\text{ohem}\star}+\alpha \|\hat{\boldsymbol{\sigma}}\|_1,
\end{equation}
where $\alpha$ is a hyper-parameter controlling the weight of the regularization term.

\section{Numerical Results}
\label{sec-simu}

\subsection{Simulation Settings and Metrics}

We consider a cellular network comprising $N_{\text{b}}=4$ BSs that simultaneously serve communication users and monitor aerial flight activities. 
The center carrier frequency is set to $f_0=2.6\ \text{GHz}$.
Each BS, positioned at the corners of a square, has a height of $\hbar_{\text{bs}}=20\ \text{m}$. 
The integrated antenna gain is $G_{\text{s}}=4$ \cite{sharma2022mimo}.
The additive noise power per receiving antenna is $P_{\text{n}}=-110\ \text{dBm}$ \cite{li2023towards}, with total noise power scaling according to the number of antennas.
The UAV RCS is randomly generated according to a Gaussian distribution with the mean 0.01 $\text{m}^2$ \cite{liu2024cooperative} and the variance 0.001. The UAV scattering coefficient is derived as the square root of its RCS \cite{huang2023joint}.

We evaluate both 2D and 3D ROIs.
Unless otherwise stated, the 2D ROI is positioned at $\hbar_{\text{roi}}=40\ \text{m}$ with dimensions of $120\ \text{m}\times 120\ \text{m}$, discretized into a $40 \times 40$ image with a voxel size of $d_0=3\ \text{m}$.
The 3D ROI has dimensions $100\ \text{m}\times 100\ \text{m}\times 80\ \text{m}$ and is discretized into a $20\times20\times16$ image with a voxel size of $d_0=5\ \text{m}$.
The imaging resolution $d_0$ is deemed sufficient for intrusion detection and trajectory planning applications \cite{ChinaMobile2024WhitePaper,3gpp2023study}.

To assess sensing performance, we employ the following five metrics:

\textbf{(1) MSE:} Measures the per-voxel difference between the predicted $\hat{\boldsymbol{\sigma}}$ and ground truth $\boldsymbol{\sigma}$ images:
\begin{equation}\label{eq-mse}
\text{MSE}=\|\hat{\boldsymbol{\sigma}}-\boldsymbol{\sigma}\|^2_2/N_{\text{v}}.
\end{equation}

\textbf{(2) Structural similarity index measure (SSIM):} Assesses the structural similarity between $\hat{\boldsymbol{\sigma}}$ and $\boldsymbol{\sigma}$ \cite{wang2024dreamer}:
\begin{equation}\label{eq-ssim}
\text{SSIM}=\frac{\left(2 \mu_{\boldsymbol{\sigma}} \mu_{\hat{\boldsymbol{\sigma}}}+c_{1}\right)\left(2 \theta_{\boldsymbol{\sigma} \hat{\boldsymbol{\sigma}}}+c_{2}\right)}{\left(\mu_{\boldsymbol{\sigma}}^{2}+\mu_{\hat{\boldsymbol{\sigma}}}^{2}+c_{1}\right)\left(\theta_{\boldsymbol{\sigma}}^{2}+\theta_{\hat{\boldsymbol{\sigma}}}^{2}+c_{2}\right)},
\end{equation}
where $\mu_{\boldsymbol{\sigma}}$ ($\mu_{\hat{\boldsymbol{\sigma}}}$) and $\theta_{\boldsymbol{\sigma}}^{2}$ ($\theta_{\hat{\boldsymbol{\sigma}}}^{2}$) are the average and variance of $\boldsymbol{\sigma}$ ($\hat{\boldsymbol{\sigma}}$), respectively.
$\theta_{\boldsymbol{\sigma} \hat{\boldsymbol{\sigma}}}$ is the covariance of $\boldsymbol{\sigma}$ and $\hat{\boldsymbol{\sigma}}$.
Constants $c_{1}$ and $c_{2}$ use MATLAB’s default settings. SSIM ranges from 0 to 1, with higher values indicating better similarity.

\textbf{(3) Optimal sub-pattern assignment (OSPA):} Evaluates target position and number estimation accuracy \cite{schuhmacher2008consistent}:
\begin{equation}\label{eq-ospa}
\text{OSPA}=\frac{1}{M_{\text{max}}}\left(\min _{\varrho \in \Pi_{M_{\text{min}}}} \sum_{m=1}^{M_{\text{min}}}\left\|\hat{\mathbf{p}}_{m}-\mathbf{p}_{\varrho(m)}\right\|_{2}+c_{3} M_{\Delta}\right).
\end{equation}
Here, $M_{\max} = \max\{M, \hat{M}\}$, $M_{\text{min}} = \min\{M, \hat{M}\}$, and $M_{\Delta} = |M - \hat{M}|$, where $\hat{M}$ is the number of detected targets in the estimated image $\hat{\boldsymbol{\sigma}}$.
$\varrho$ is one element in the set $\Pi_{M_{\text{min}}}$, which represents all possible permutations on $\{1, 2, \ldots, M_{\text{min}}\}$.
$\mathbf{p}_m$ and $\hat{\mathbf{p}}_m$ denote true and estimated target locations.
The penalty term $c_{3} M_{\Delta}$ with constant $c_3=1$ is considered to measure the target number estimation error.

\textbf{(4) DR:} Represents the proportion of correctly identified targets in the reconstructed image.

\textbf{(5) FAR:} Indicates the proportion of falsely detected targets that do not exist in the ground truth.

\subsection{On-Grid Simulation Results}
\label{sec-simu-on}

This subsection utilizes Algorithm \ref{ag1} to discuss the influences of system configurations on sensing performance under on-grid conditions.

\subsubsection{Sensing Performance with Varying Antenna Numbers, BS Distances, and Transmit Powers}
\label{sec-simu-on-1}

\begin{figure}
    \centering
    \includegraphics[width=0.85\linewidth]{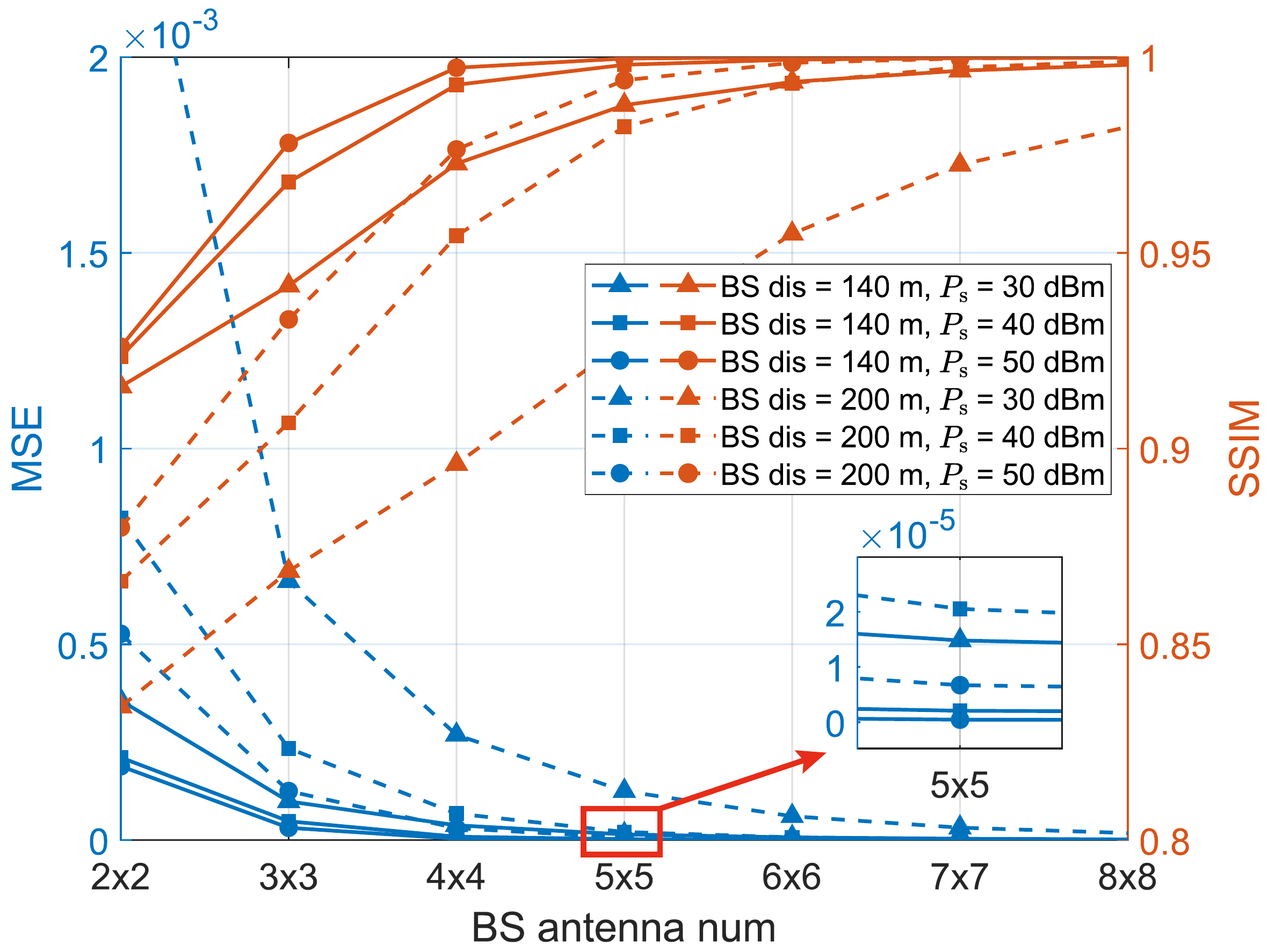}
    \captionsetup{font=footnotesize}
    \caption{Sensing performance under on-grid conditions with varying antenna numbers, BS distances, and transmit powers.}
    \label{fig-result-on-1}
\end{figure}

We evaluate the sensing performance of the proposed imaging-based surveillance method using the MSE and SSIM metrics. The results in Fig.~\ref{fig-result-on-1} represent the average of 10,000 Monte Carlo simulations. The number of UAVs is set to $M=6$.
As the number of BS antennas increases, more CSI measurements can be leveraged for reconstructing low-altitude images. Consequently, the MSE decreases while the SSIM improves, indicating enhanced sensing performance.
However, increasing the distance between BSs can degrade sensing performance due to stronger channel correlations among antennas, which enlarge the condition number of the sensing matrix. Nevertheless, these negative effects can be partially mitigated by increasing the sensing signal power or deploying larger transceiving antenna arrays. Therefore, wireless imaging-based surveillance of low-altitude airspace remains effective with proper system configurations.

\subsubsection{Sensing Performance with Different Sensing Modes and Voxel Sizes}
\label{sec-simu-on-2}

\begin{table}[t]
    \renewcommand{\arraystretch}{1.3}
    \centering
    \fontsize{8}{8}\selectfont
    \captionsetup{font=small}
    \caption{Sensing performance under on-grid conditions with varying sensing modes and voxel sizes.}\label{tab-on-grid}
    \begin{threeparttable}
\begin{tabular}{p{1cm}<{\centering}cp{0.85cm}<{\centering}p{0.85cm}<{\centering}p{0.85cm}<{\centering}p{0.85cm}<{\centering}p{0.85cm}<{\centering}}
            \specialrule{1pt}{0pt}{-1pt}\xrowht{15pt}
            & \diagbox[width=3.8em]{\hspace{-0.15cm}mode}{\hspace{0.1cm}$d_0$} & 1 m & 2 m & 3 m & 4 m & 5 m \\
            \hline
            \multirow{3}{*}{\shortstack{MSE\\($\times10^{-4}$)}} & A & \textbf{1.8598} & \textbf{0.3293} & \textbf{0.0574} & \textbf{0.0444} & \textbf{0.0429} \\
            & B & 60.741 & 2.0121 & 0.7762 & 0.4432 & 0.2236 \\
            & C & 466.72 & 21.370 & 13.519 & 7.9370 & 4.6723 \\
            \multirow{3}{*}{DR} & A & \textbf{40.78\%} & \textbf{83.22\%} & \textbf{96.47\%} & \textbf{98.99\%} & \textbf{99.78\%} \\
            & B & 24.07\% & 63.63\% & 86.34\% & 95.27\% & 98.02\% \\
            & C & 0.32\% & 2.25\% & 8.67\% & 22.58\% & 43.24\% \\
            \multirow{3}{*}{FAR} & A & \textbf{70.92\%} & \textbf{34.04\%} & \textbf{12.60\%} & \textbf{5.37\%} & \textbf{1.18\%} \\
            & B & 81.63\% & 54.15\% & 31.76\% & 18.73\% & 8.88\% \\
            & C & 98.77\% & 93.26\% & 78.99\% & 60.44\% & 44.09\% \\
            \specialrule{1pt}{0pt}{0pt}
        \end{tabular}
    \end{threeparttable}
\end{table}

We analyze the sensing performance of different sensing modes with varying imaging resolution $d_0$, setting $N_0=4$, and $P_{\text{s}}=40\ \text{dBm}$. The BS distance is fixed at 140 m.
Table \ref{tab-on-grid} presents the simulation results. Mode ``A'' refers to the proposed joint monostatic and multi-static sensing scheme, where all BSs receive sensing signals transmitted by any of them. 
In Mode ``B'', signals transmitted by one BS can only be received by other BSs \cite{liu2024cooperative}, while in Mode ``C'', each BS can only receive signals transmitted by itself \cite{tang2024cooperative}.
Among the three, Mode ``A'' achieves the best sensing performance by aggregating the highest number of CSI measurements for image reconstruction. This demonstrates that integrating monostatic and multi-static sensing capabilities is essential for high-performance imaging. 
Additionally, all three modes exhibit improved imaging accuracy with increasing $d_0$.

\begin{figure}[t]
    \centering
    \includegraphics[width=\linewidth]{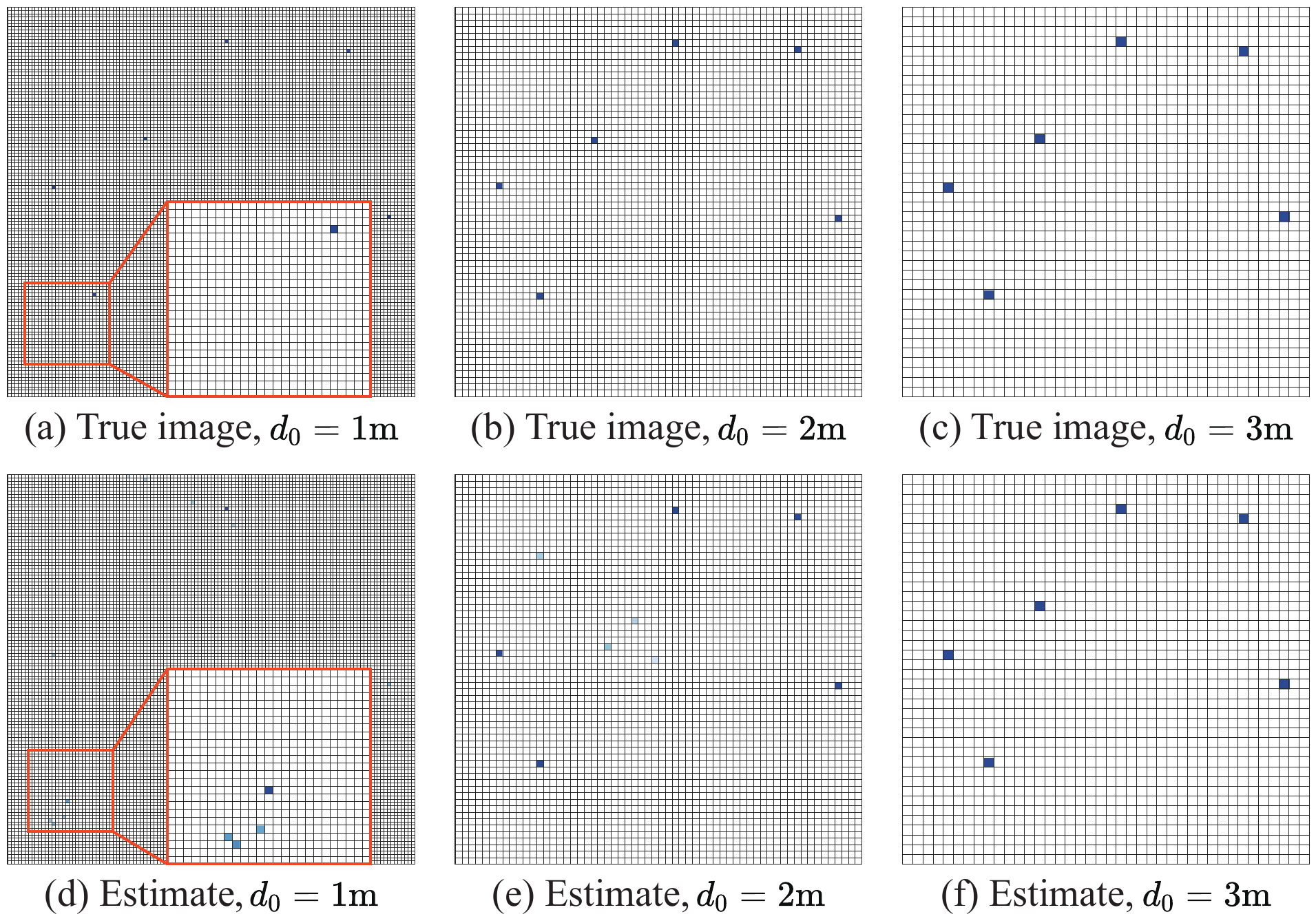}
    \captionsetup{font=footnotesize}
    \caption{Imaging results under on-grid conditions for different voxel sizes.}
    \label{fig-result-on-2}
\end{figure}

Fig.~\ref{fig-result-on-2} illustrates true and predicted images under Mode ``A.'' Perfect image reconstruction is achieved at $d_0=3\ \text{m}$, as seen in Fig.~\ref{fig-result-on-2}(f).
However, reducing $d_0$ increases channel correlations among voxels, making them harder to be distinguished, as analyzed in Sec. \ref{sec-psf} using the PSF.
For example, one target is lost at $d_0=2\ \text{m}$, as seen in Fig.~\ref{fig-result-on-2}(e), while nearly no targets are accurately detected at $d_0=1\ \text{m}$, as shown in Fig.~\ref{fig-result-on-2}(d). This indicates that an excessively fine imaging resolution may degrade performance due to increased correlations among voxels' steering vectors.
In summary, the proposed joint monostatic and multi-static sensing scheme provides superior imaging performance, and selecting an appropriate imaging resolution is critical to ensure accurate surveillance.

\subsubsection{Sensing Performance with Varying Subcarrier Numbers, Bandwidths, and Antenna Spacings}
\label{sec-simu-on-3}

\begin{figure}[t]
    \centering
    \includegraphics[width=0.85\linewidth]{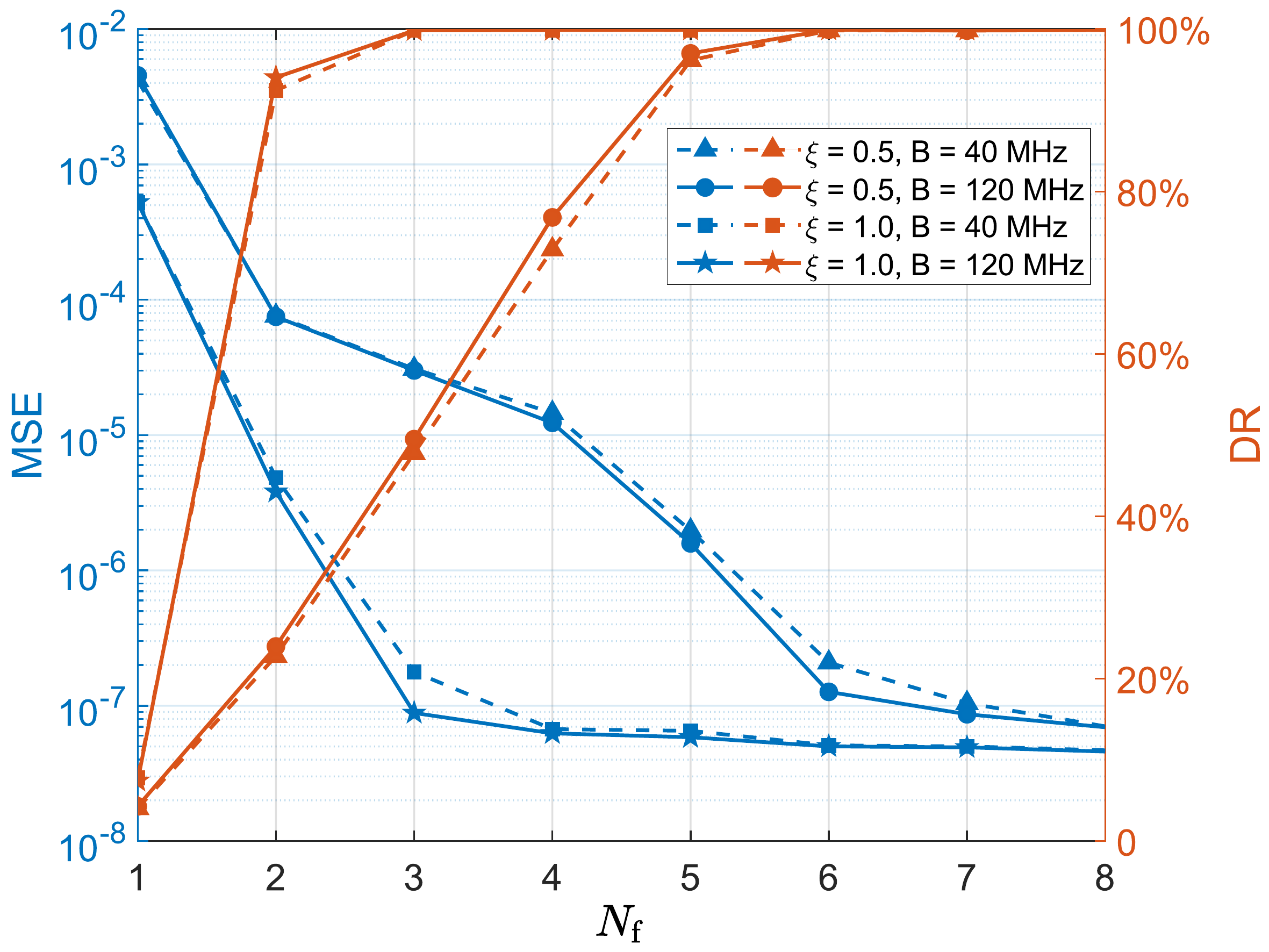}
    \captionsetup{font=footnotesize}
    \caption{Sensing performance of the 3D ROI with varying subcarrier numbers, bandwidths, and antenna spacings.}
    \label{fig-result-on-3}
    \vspace{-0.2cm}
\end{figure}

We investigate the impact of subcarrier number $N_{\text{f}}$, bandwidth $B$, and antenna spacing $(\lambda_0/2)\cdot\xi$ on the 3D imaging performance. 
The ROI center is at $\hbar_{\text{roi}}=80\ \text{m}$, and the number of UAVs is increased to $M=24$.
Since a larger $N_{\text{f}}$ provides more CSI measurements, we employ UPAs with $2\times 2$ antennas to derive the simulation results presented in Fig.~\ref{fig-result-on-3}. The total sensing signal power is set to $P_{\text{s}}=40\ \text{dBm}$. Although increasing $N_{\text{f}}$ reduces the power allocated per subcarrier, it leads to a lower MSE and a higher DR. This suggests that utilizing multiple subcarriers is advantageous for imaging, as it enables the acquisition of richer environmental information.
However, similar to the observations in Fig.~\ref{fig-psf-3}, a continuous increase in $N_{\text{f}}$ does not necessarily enhance imaging accuracy when $B$ is limited. Additionally, larger bandwidths can further improve sensing performance. Finally, deploying sparse arrays effectively reduces the MSE and increases the DR, as the enlarged imaging aperture facilitates the collection of more environmental details and enhances spatial resolution.

\subsubsection{Tradeoff Between Communication and Sensing Performance}
\label{sec-simu-on-4}

\begin{figure}[t]
    \centering
    \includegraphics[width=0.85\linewidth]{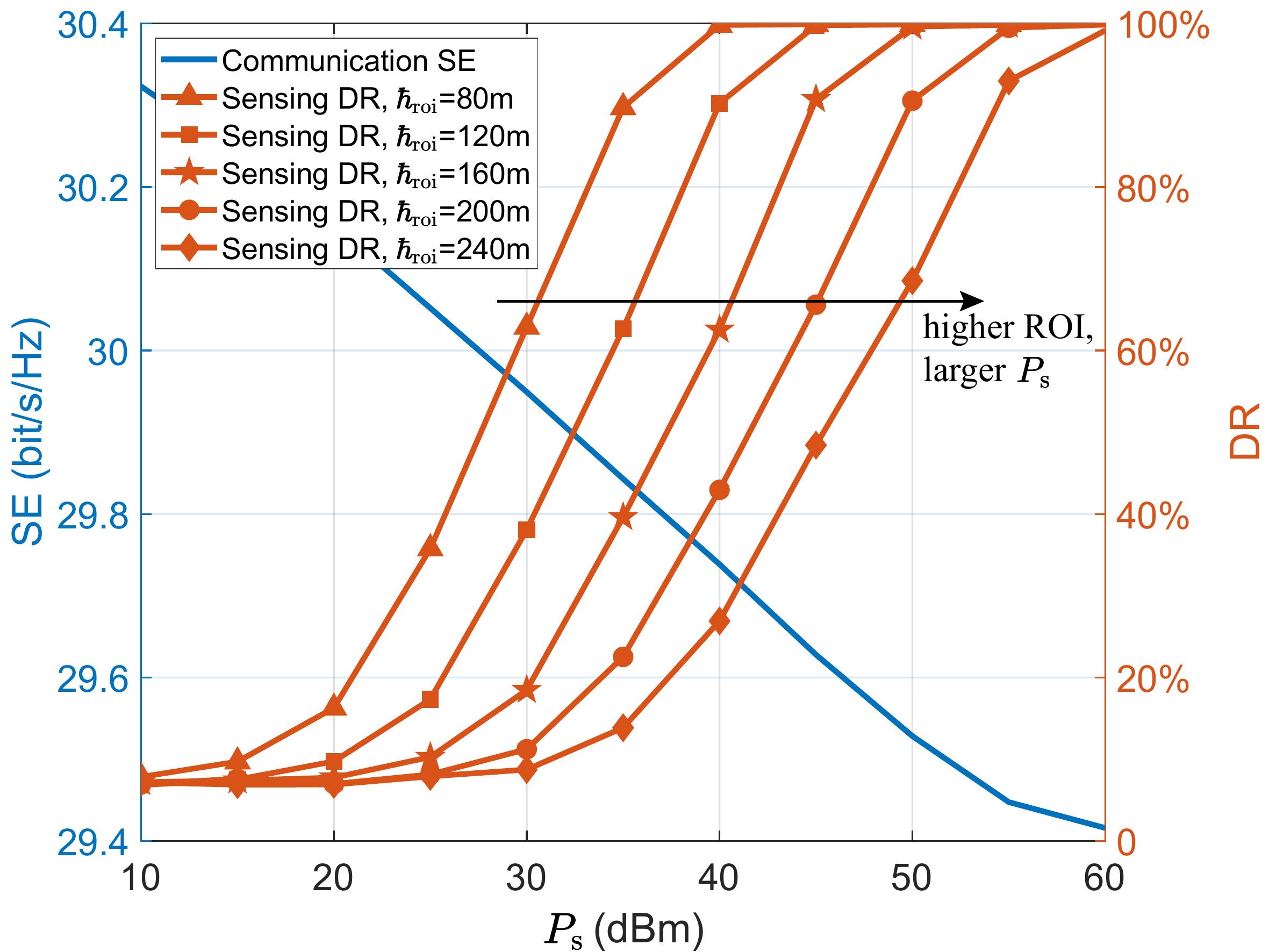}
    \captionsetup{font=footnotesize}
    \caption{Tradeoff between communication and sensing performances.}
    \label{fig-result-on-4}
\end{figure}

We consider an ISAC system in which one aerial image is captured per frame, utilizing $N_{\text{s}} = N_0^2 = 4$ ISAC symbol intervals out of a total of 140 symbols per frame.
The sensing signal power is set to $P_{\text{s}} = 0$ during pure communication intervals.
During the ISAC period, BSs simultaneously direct beams toward both the ROI and the communication users.
The beamforming matrix $\mathbf{W}_{n_{\text{b}}}$ is designed according to \cite{li2023towards,wang2024heterogeneous} to minimize power leakage across different beams. This ensures that the multi-user interference term in \eqref{eq-commun-signal-model} remains negligible.
Similar to Sec. \ref{sec-simu-on-3}, we employ $N_{\text{f}}=6$ subcarriers within a bandwidth of $B=40\ \text{MHz}$ to construct the 3D ROI image. The antenna spacing is set to $\lambda_0/2$.
For simplicity, we analyze a single communication user positioned at $[0, 0, 0]^{\text{T}}$.

Fig.~\ref{fig-result-on-4} presents the simulation results for a total transmission power of $P_{\text{t}}=60\ \text{dBm}$. 
The DR increases with $P_{\text{s}}$ and reaches 100\% when $P_{\text{s}}=40\ \text{dBm}$ and the ROI center height is $\hbar_{\text{roi}}=80\ \text{m}$.
As $\hbar_{\text{roi}}$ increases, achieving perfect target detection requires higher sensing signal power. This increase in $P_{\text{s}}$ slightly degrades the communication SE. However, the degradation remains minimal because the imaging function occupies only a small portion of the available communication resources.
Our simulation results show that when $P_{\text{s}} = 40\ \text{dBm}$, the desired communication signal power received by the user is approximately 24 dB higher than the interference caused by the sensing signals.
As a result, the impact of sensing signals on communication performance is negligible. This is mainly due to the low UAV RCS and the long signal transmission distance.
Therefore, the SE is primarily determined by the communication signal power $P_{\text{c}}$, highlighting the need to balance $P_{\text{s}}$ and $P_{\text{c}}$ to optimize ISAC performance.

\subsection{Off-Grid Simulation Results}
\label{sec-simu-off}

This section evaluates the proposed imaging algorithms from Sec. \ref{sec-imaging-algorithm} under off-grid conditions. The BS distance is set to 140 m, the antenna array size is $5 \times 5$, and the transmit power is $P_{\text{t}}=40 \ \text{dBm}$.\footnote{
This configuration results in a $6250\times 1600$ matrix per frequency, requiring 3D partial derivatives for the Taylor expansion method \cite{yang2012off,wei2022off}. Moreover, running the ANM-based algorithm \cite{li2024atomic,gao2024robust} exceeds 32 GB of memory.}
The DNN used in this subsection consists of six residual blocks. The number of channels in the convolutional layers is set to $[64, 128, 128, 128, 64, 32]$.
Network parameters are optimized via RMSprop on an Nvidia A100 GPU using PyTorch, trained for 200 epochs with an initial learning rate of 0.001, halving if the loss stagnates for five epochs. The dataset includes 100,000 MATLAB-generated images (10\% for validation), with an additional 10,000 test images.
The average UAV count is 6 for 2D and 12 for 3D ROIs, as detailed in Secs. \ref{sec-simu-off-1} to \ref{sec-simu-off-4} and Sec. \ref{sec-simu-off-5}, respectively.

\subsubsection{Sensing Performance Comparison Across Different Algorithms}
\label{sec-simu-off-1}

\begin{table*}[t]
    \renewcommand{\arraystretch}{1.3}
    \centering
    \fontsize{8}{8}\selectfont
    \captionsetup{font=small}
    \caption{Sensing results of different algorithms for off-grid targets.}\label{tab-off-grid-1}
    \begin{threeparttable}
        \begin{tabular}{ccccccc}
            \specialrule{1pt}{0pt}{-1pt}\xrowht{10pt}
            Methods & MSE & SSIM & OSPA & DR & FAR & Run Time (ms) \\
            \hline
            SP & 0.0067 & 0.6909 & 27.7465 & 46.52\% & 69.41\% & 52.05 (CPU) \\
            $\mathbf{A}^{\text{H}}\mathbf{y}$ & 0.0308 & 0.0534 & 72.5825 & 45.56\% & 85.99\% & 0.08 (GPU) \\
            DNN$\sphericalangle\mathbf{y}$ & 0.0037 & 0.6895 & 50 & 0 & 0 & 11.41 (GPU) \\
            Model+DNN$\sphericalangle$SP & 0.0033 & 0.7778 & 29.2561 & 35.06\% & 15.28\% & 65.07 (GPU) \\
            Model+DNN$\sphericalangle\mathbf{A}^{\text{H}}\mathbf{y}$ & \textbf{0.0009} & \textbf{0.9186} & \textbf{7.5957} & \textbf{86.24\%} & \textbf{3.29\%} & 13.35 (GPU) \\
            \specialrule{1pt}{0pt}{0pt}
        \end{tabular}
    \end{threeparttable}
\end{table*}

\begin{figure}[t]
    \centering
    \includegraphics[width=\linewidth]{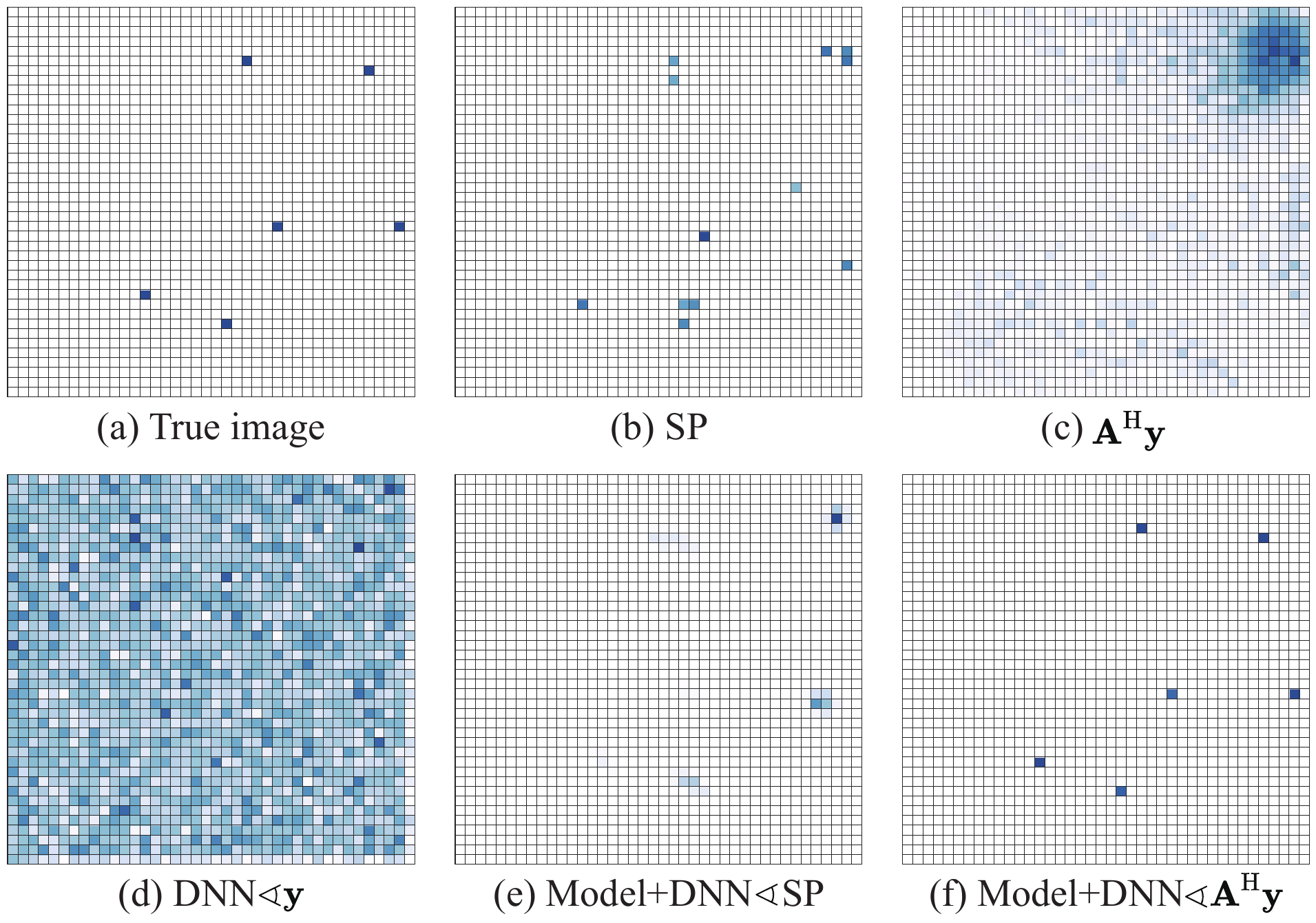}
    \captionsetup{font=footnotesize}
    \caption{Imaging results of different methods under off-grid conditions.}
    \label{fig-result-off-1}
\end{figure}

We compare the sensing performance of various imaging methods for off-grid UAV positions, including: the SP algorithm with on-grid sensing matrix, the intermediate sensing result $\mathbf{A}^{\text{H}}\mathbf{y}$, and a DNN trained using the CSI measurement $\mathbf{y}$ as the input (DNN$\sphericalangle\mathbf{y}$).
The proposed physics-embedded learning method in Sec. \ref{sec-algo-off-grid} is validated by training DNNs using the outputs of the SP algorithm (Model+DNN$\sphericalangle$SP) and using $\mathbf{A}^{\text{H}}\mathbf{y}$ (Model+DNN$\sphericalangle\mathbf{A}^{\text{H}}\mathbf{y}$).
All DNN models are designed with a comparable number of trainable parameters and are trained using the same strategy.
According to the simulation results in Table \ref{tab-off-grid-1}, the proposed method ``Model+DNN$\sphericalangle\mathbf{A}^{\text{H}}\mathbf{y}$'' achieves the best performance across all evaluation metrics.
Moreover, the execution time of the proposed method shows that it can achieve nearly real-time inference, primarily due to the high computational efficiency of the GPU.
In contrast, the iteration-based SP algorithm executed on the CPU requires the longest computation time.

Fig.~\ref{fig-result-off-1} further illustrates the imaging results produced by these methods. The SP algorithm struggles to correctly identify target voxels, often producing false detections due to model mismatch. The intermediate result $\mathbf{A}^{\text{H}}\mathbf{y}$ and the DNN trained with $\mathbf{y}$ exhibit substantial visual noise, capturing little to no meaningful image information.
While the ``Model+DNN$\sphericalangle$SP'' method partially detects targets, it frequently misplaces them due to distortions introduced by the SP algorithm. In contrast, the proposed ``Model+DNN$\sphericalangle\mathbf{A}^{\text{H}}\mathbf{y}$'' method achieves precise image reconstruction and accurate non-zero voxel detection, demonstrating its superior capability in off-grid sensing.

\subsubsection{Sensing Performance Comparison Across Different DNN Structures and Loss Functions}
\label{sec-simu-off-2}

\begin{figure*}
\begin{minipage}{0.65\linewidth}
    \renewcommand{\arraystretch}{1.3}
    \centering
    \fontsize{8}{8}\selectfont
    \captionsetup{font=small}
    \captionof{table}{Sensing performance across different DNN structures and loss functions.}\label{tab-off-grid-2}
    \begin{threeparttable}
\begin{tabular}{ccccp{0.7cm}<{\centering}p{0.7cm}<{\centering}p{0.85cm}<{\centering}p{0.85cm}<{\centering}p{0.85cm}<{\centering}}
            \specialrule{1pt}{0pt}{-1pt}\xrowht{10pt}
            Network & DNN Structure & Loss Fun. & $\alpha$ & MSE & SSIM & OSPA & DR & FAR \\
            \hline
            Net-1 & Dn-CNN & MSE & 0 & 0.0021 & 0.8117 & 21.1022 & 73.99\% & 3.97\% \\
            Net-2 & Dn-resCNN & MSE & 0 & 0.0021 & 0.8200 & 20.6430 & 75.19\% & 4.09\% \\
            Net-3 & CNN & MSE & 0 & 0.0019 & 0.8352 & 19.1941 & 76.25\% & 3.91\% \\
            Net-4 & CNN & MSE & 1 & 0.0037 & 0.7699 & 50      & 0      & 0      \\
            Net-5 & resCNN & MSE & 0 & 0.0017 & 0.8457 & 17.5013 & 78.29\% & 3.93\% \\
            Net-6 & resCNN & MSE & 1 & 0.0037 & 0.7699 & 50      & 0      & 0      \\
            Net-7 & resCNN & OHEM-1 & 0 & 0.0018 & 0.8658 & 16.7250 & 76.24\% & 3.95\% \\
            Net-8 & resCNN & OHEM-2 & 0 & 0.0012 & 0.9156 &  8.7960 & 86.48\% & 4.01\% \\
            Net-9 & resCNN & OHEM-1 & 1 & \textbf{0.0009} & \textbf{0.9186} & \textbf{7.5957} & 86.24\% & 3.29\% \\
            Net-10 & resCNN & OHEM-2 & 1 & 0.0029 & 0.8546 & 29.9405 & \textbf{97.55\%} & \textbf{3.22\%} \\
            \specialrule{1pt}{0pt}{0pt}
        \end{tabular}
    \end{threeparttable}
\end{minipage}
\
\begin{minipage}{0.34\linewidth}
    \centering
    \includegraphics[width=0.95\linewidth]{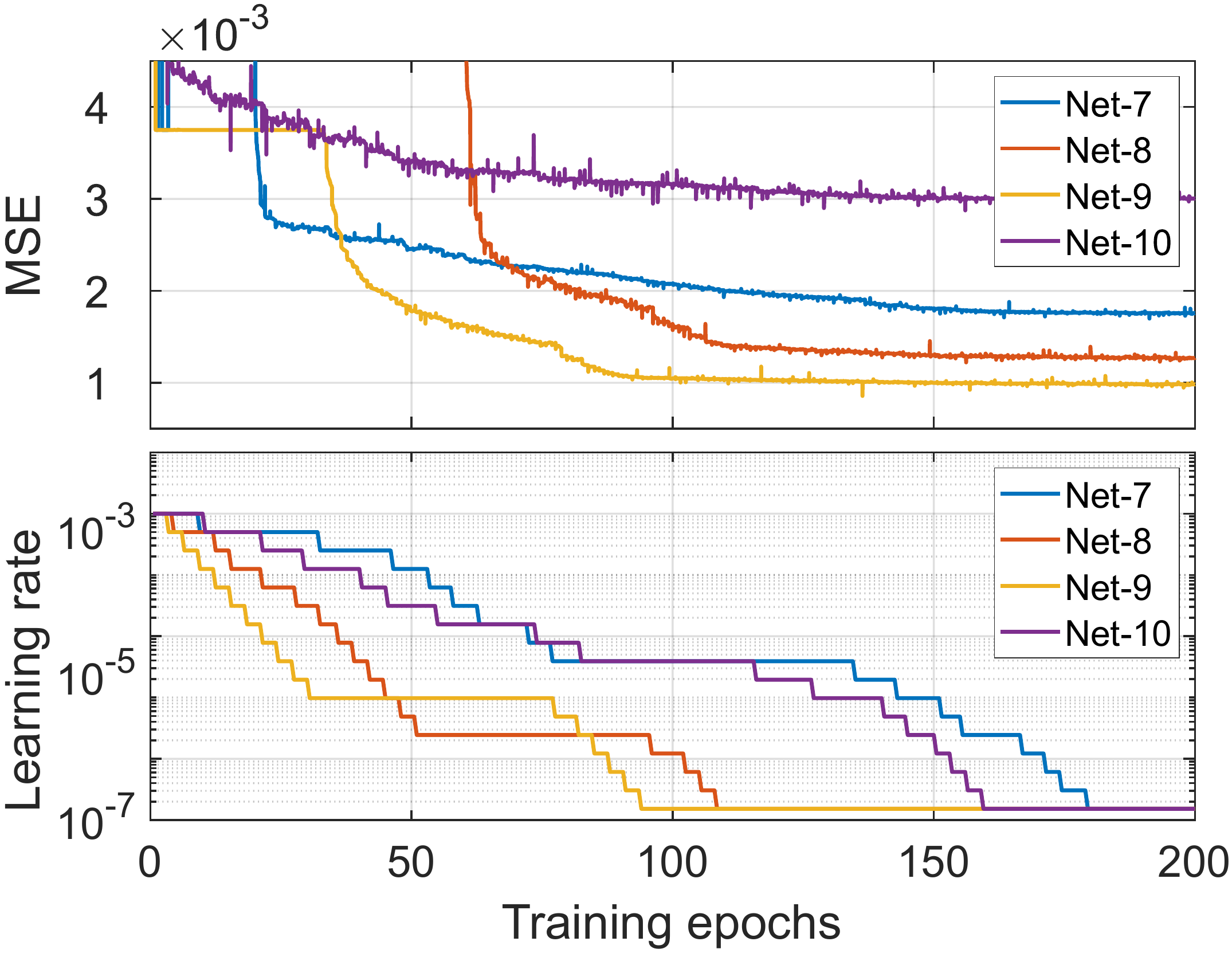}
    \captionsetup{font=footnotesize}
    \vspace{-0.2cm}
    \caption{MSE and learning rate over training epochs.}
    \label{fig-result-off-2}
\end{minipage}
\vspace{-0.15cm}
\end{figure*}

This section examines how different DNN architectures and loss functions affect training performance. The ``resCNN'' model employs the residual structure illustrated in Fig.~\ref{fig-algorithm-flow}(b). Additionally, an image denoising approach is tested, where the DNN is trained to estimate the noise component $\boldsymbol{\sigma}_{\text{pri}}-\boldsymbol{\sigma}$ (Dn-CNN and Dn-resCNN).
The simulation results in Table \ref{tab-off-grid-2} reveal that the Dn-CNN does not perform well in this imaging scenario, indicating that the noise in $\boldsymbol{\sigma}_{\text{pri}}$ lacks a structured probability distribution that can be effectively learned by the DNN. Introducing a residual structure improves sensing performance, as seen in the comparison of different architectures. However, when sparse regularization is applied, as in Net-4 and Net-6, the networks tend to output all-zero images, leading to a failure in detecting low-altitude targets.

Training the DNNs with the proposed OHEM loss functions significantly enhances performance across all test metrics, particularly when comparing Net-5 and Net-8. The inclusion of a sparse regularization term further enhances the network’s effectiveness, as demonstrated by Net-9 and Net-10. The MSEs and learning rates of the last four DNNs during training, shown in Fig.~\ref{fig-result-off-2}, highlight the crucial role of the loss function in determining DNN performance.
Among these models, Net-9, trained with the OHEM-1 loss function and $\alpha=1$, achieves the lowest MSE and OSPA while obtaining the highest SSIM. In contrast, Net-10 achieves the best DR, identifying 97.55\% of targets in the test dataset while maintaining a low FAR of 3.22\%. This suggests that while Net-9 is optimal for accuracy-focused applications, Net-10 is preferable when maximizing the DR and minimizing missed targets. 

\begin{figure*}
\centering
\captionsetup{font=footnotesize}
\begin{subfigure}[b]{0.38\linewidth}
\centering
\includegraphics[width=\linewidth]{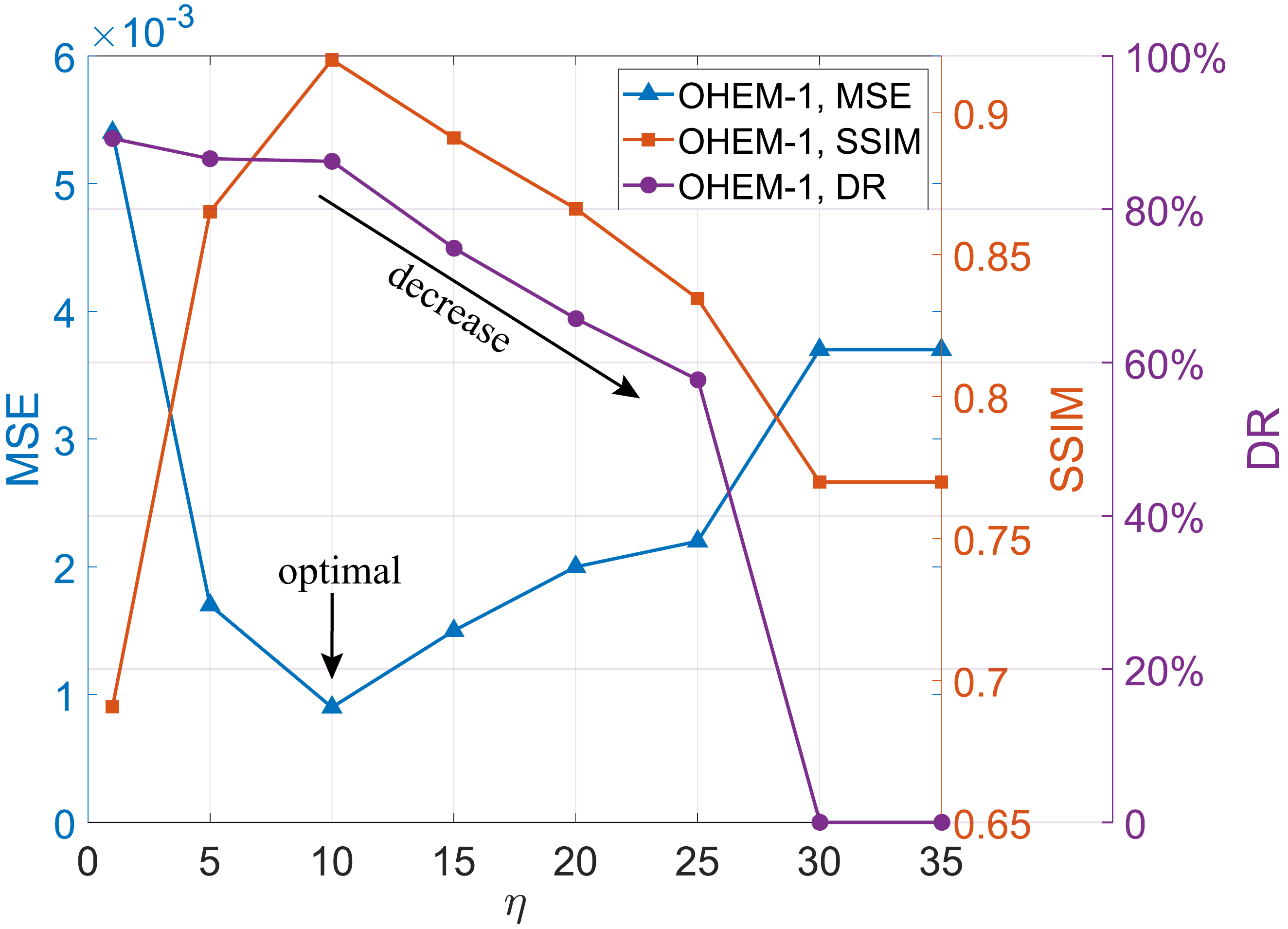}
\caption{OHEM-1}
\label{fig-result-off-3-ohem1}
\end{subfigure}
\quad\quad
\begin{subfigure}[b]{0.38\linewidth}
\centering
\includegraphics[width=\linewidth]{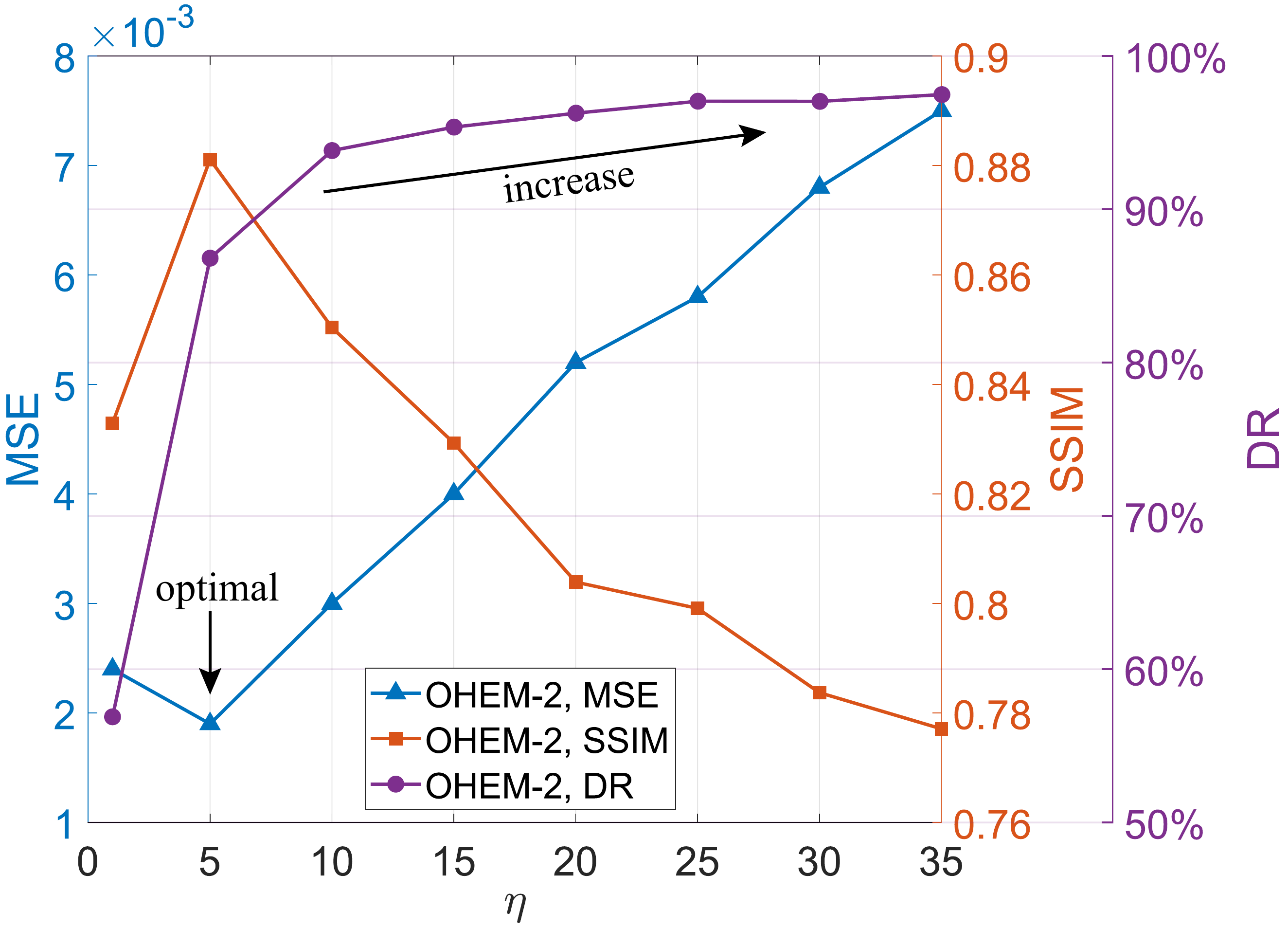}
\caption{OHEM-2}
\label{fig-result-off-3-ohem2}
\end{subfigure}
\caption{Sensing performance variation with different ratios ($\eta$) of selected passive to positive sample numbers.}
\label{fig-result-off-3-ohem}
\vspace{-0.27cm}
\end{figure*}

\subsubsection{Influence of $\eta$ on OHEM-Based DNN Training}
\label{sec-simu-off-3}

This section examines how the negative sample ratio $\eta$ affects sensing performance by training a series of DNNs. The performance of the networks trained with OHEM-1 and OHEM-2 loss functions is shown in Fig.~\ref{fig-result-off-3-ohem1} and Fig.~\ref{fig-result-off-3-ohem2}, respectively.

For OHEM-1, the three evaluation metrics, namely MSE, SSIM, and DR, exhibit different trends as $\eta$ varies. When $\eta$ is small, the network tends to generate a large number of non-zero voxels in the reconstructed images, which leads to inconsistencies with the true labels. As a result, the MSE remains high and the SSIM is low when 
$\eta=1$, indicating poor sensing performance, although the DR reaches its peak. As $\eta$ increases, the number of selected negative samples $N_{\text{neg}}$ grows, enhancing the loss function’s focus on negative samples, as described in \eqref{eq-loss-ohem}. When $\eta=10$, a balance is achieved between attention to positive and negative samples, yielding the lowest MSE and highest SSIM while maintaining a DR only slightly lower than its peak value. Beyond this point, further increases in $\eta$ cause the network to output images with more zero-value voxels, leading to a degradation in DR and SSIM. When $\eta \ge 30$, the network outputs all-zero images, capturing no target information.

For OHEM-2, the trends in MSE and SSIM are similar to those observed in OHEM-1. However, $\eta=5$ achieves the best tradeoff between positive and negative samples in this case. Unlike OHEM-1, the DR continuously increases with growing $\eta$ due to the weakened contribution of each passive sample, as reflected in the second term of \eqref{eq-loss-ohem}. The DR saturates at approximately 98\% when $\eta\ge 25$, which marks the most significant difference in training behavior between OHEM-1 and OHEM-2.

These results underscore the critical role of $\eta$ in optimizing sensing performance. The optimal $\eta$ depends on the specific loss function and application, requiring careful selection to balance detection accuracy and robustness.

\subsubsection{Evaluation of the Adaptability of the Proposed DNN to Different Environments}
\label{sec-simu-off-4}

\begin{table}[t]
    \renewcommand{\arraystretch}{1.3}
    \centering
    \fontsize{8}{8}\selectfont
    \captionsetup{font=small}
    \caption{OSPA performance across varying environments (units: OSPA in meter, $P_{\text{n}}$ in dBm).}\label{tab-off-grid-4}
    \begin{threeparttable}
        \begin{tabular}{cccccc}
            \specialrule{1pt}{0pt}{-1pt}\xrowht{10pt}
            Dataset & 2 & 3 & 4 & 5 & 6 \\
            \hline
            UAV number & 6 & 6 & 6 & 4 & 8 \\
            $P_{\text{n}}$ & $-120$ & $-110$ & $-100$ & / & / \\
            OSPA-A & 9.6810 & 10.2537 & 12.1318 & 6.9136 & 11.3106 \\
            OSPA-B & 9.0680 & 10.5033 & 12.3616 & 6.5370 & 11.2219 \\
            \specialrule{1pt}{0pt}{0pt}
        \end{tabular}
    \end{threeparttable}
\end{table}

To assess the adaptability of the proposed DNN, six new datasets are generated for training and testing. The first dataset is constructed with a ROI containing six UAVs, with the additive noise power set to $P_{\text{n}}=0$.
The parameter settings for the remaining five datasets are provided in Table \ref{tab-off-grid-4}.
Using the OHEM-1 loss function with $\alpha=1$, the proposed DNN is trained on the first dataset and achieves an OSPA of 9.6159 m on the corresponding test set. Without additional retraining, the same DNN is directly tested on the other five datasets, and the OSPA results are listed in the ``OSPA-A'' row of Table \ref{tab-off-grid-4}. Additionally, five separate DNNs are trained using the respective datasets and evaluated using their corresponding test data, with the results presented in the ``OSPA-B'' row of Table \ref{tab-off-grid-4}.
The comparison reveals that the DNN trained on the first dataset performs well on unseen datasets with varying noise powers and UAV numbers, achieving results comparable to those of DNNs specifically trained on each dataset. This demonstrates that the proposed method exhibits strong adaptability to diverse environments. The enhanced generalization capability is likely attributed to the embedded physics-based information in the input primary result, as formulated in \eqref{eq-primary}.
Furthermore, sensing performance degrades with the noise power and UAV number increasing, owing to the corresponding higher difficulty in low-altitude imaging.

\subsubsection{3D ROI Imaging in Sionna}
\label{sec-simu-off-5}

\begin{figure}[t]
    \centering
    \includegraphics[width=0.77\linewidth]{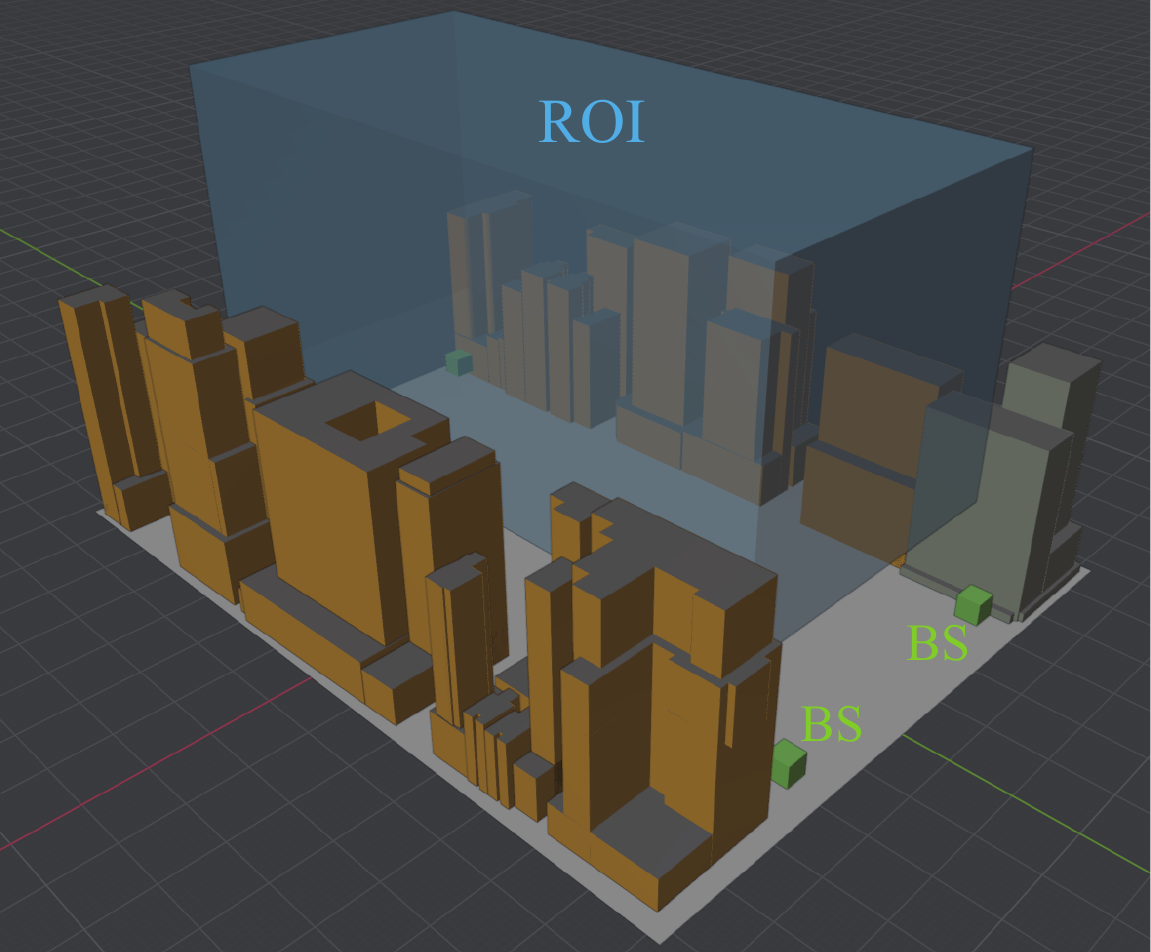}
    \captionsetup{font=footnotesize}
    \caption{The simulation scenario of urban canyon environments in Sionna.}
    \label{fig-sionna-scenario}
\end{figure}

This subsection employs Sionna, a widely used ray tracing channel emulator in both academic and industrial fields, to generate the sensing channels required by the proposed algorithms.
Specifically, we use a real-world road and its surroundings in Hong Kong to simulate urban canyon environments. The data is downloaded from OpenStreetMap, processed in Blender, and then used by Sionna.
The scenario is depicted in Fig.~\ref{fig-sionna-scenario}, where the canyon is 50 m wide and 130 m long, and the buildings are approximately 60 m high.
Four BSs are mounted at the four corners of the canyon at an altitude of $\hbar_{\text{bs}}=10\ \text{m}$.
The 3D ROI has dimensions of $40\ \text{m}\times 120\ \text{m}\times 40\ \text{m}$ and is discretized into an $8\times24\times8$ image with a voxel size of $d_0=5\ \text{m}$.
The number of UAVs in the ROI is set to $M=8$.
We use 3D convolutional kernels to generate 3D images of the ROI.

\begin{figure}[t]
    \centering
    \includegraphics[width=0.75\linewidth]{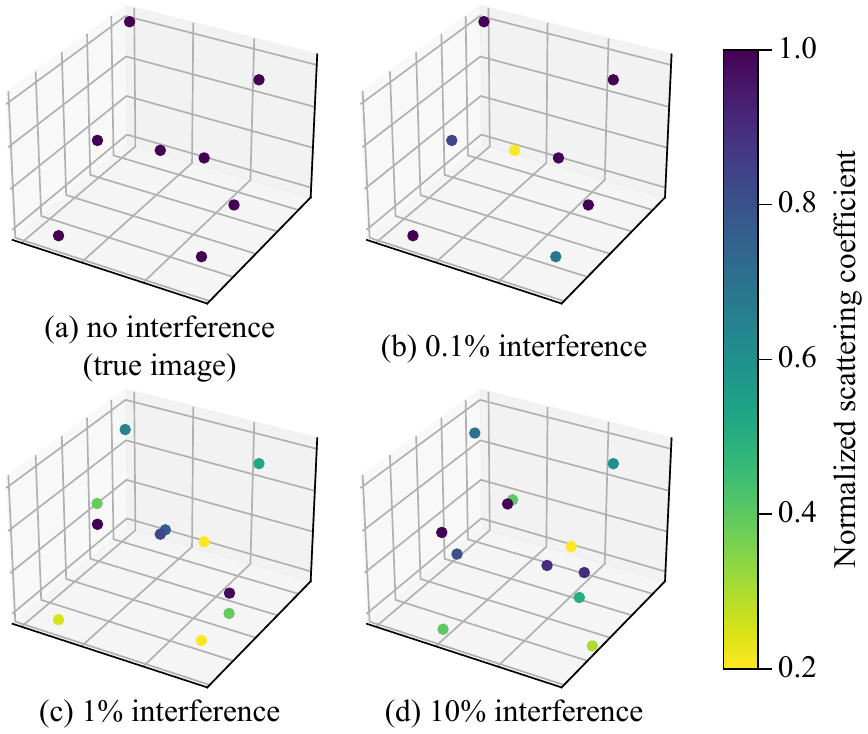}
    \captionsetup{font=footnotesize}
    \caption{3D imaging results for different residual interference ratios.}
    \label{fig-result-off-5}
\end{figure}

Considering the interference caused by surrounding buildings, dedicated beamformers oriented toward the ROI \cite{li2023towards} and calibration processes \cite{li2024radio} should be employed to suppress interference.
Table \ref{tab-off-grid-5} presents the sensing performance with varying levels of residual interference, where increasing interference results in degraded performance across all evaluated metrics.
Specifically, the DR exceeds 80\% with 0.1\% residual interference but can decrease by more than 25\% when no interference suppression techniques are applied.
Fig.~\ref{fig-result-off-5} illustrates the 3D imaging results under various interference levels. Nearly perfect image formation is achieved in Fig.~\ref{fig-result-off-5}(a) when no interference is present.
In Fig.~\ref{fig-result-off-5}(b), the DNN successfully detects all UAVs under 0.1\% residual interference, although the estimated scattering coefficients may deviate from the ground truth.
With higher levels of interference, detection errors increase and falsely detected targets appear, as shown in Fig.~\ref{fig-result-off-5}(c) and Fig.~\ref{fig-result-off-5}(d).
These results demonstrate that interference suppression using beamforming and background removal is critical to the proposed algorithms. Effective suppression enables high sensing performance.

\begin{table}[t]
    \renewcommand{\arraystretch}{1.3}
    \centering
    \fontsize{8}{8}\selectfont
    \captionsetup{font=small}
    \caption{3D sensing performance in Sionna vs. ratio of residual interference.}\label{tab-off-grid-5}
    \begin{threeparttable}
        \begin{tabular}{m{2cm}<{\centering}ccccc}
            \specialrule{1pt}{0pt}{-1pt}\xrowht{10pt}
            Ratio of residual interference & 0 & 0.1\% & 1\% & 10\% & 100\% \\
            \hline
            MSE & 0.0034 & 0.0034 & 0.0035 & 0.0035 & 0.0041 \\
            SSIM & 0.9006 & 0.7695 & 0.7437 & 0.6653 & 0.2540 \\
            DR & 83.37\% & 80.35\% & 77.40\% & 69.88\% & 57.85\% \\
            FAR & 3.57\% & 3.80\% & 4.03\% & 3.65\% & 4.10\% \\
            \specialrule{1pt}{0pt}{0pt}
        \end{tabular}
    \end{threeparttable}
\end{table}

\section{Conclusion and Future Research Directions}
\label{sec-conclusion}

This study investigates flight activity surveillance in emerging LAE scenarios by leveraging existing ISAC cellular networks.
A CS-based imaging problem is formulated under on-grid conditions, with the PSF assessing the system’s sensing capabilities.
To address the challenges posed by off-grid errors, a physics-embedded learning method is introduced, combining primary results obtained from the
traditional sensing matrix with refinements via DL techniques. The proposed approach enhances DNN training through OHEM-based loss functions, improving DRs.
Simulation results validate the effectiveness of the proposed imaging-based surveillance framework. The physics-embedded learning method accurately reconstructs both 2D and 3D images in off-grid scenarios, outperforming conventional CS-based algorithms. These findings highlight the potential of ISAC-based imaging for high-precision, real-time low-altitude surveillance, offering a scalable solution for future airspace monitoring applications.

\bl{Our future work may include optimizing the algorithm design to enhance DRs in complex environments and reduce the training overhead of the DNN.
In addition, dynamic imaging through UAV tracking across successive time instants is anticipated to improve the quality of low-altitude surveillance.
Validation through field trials conducted in live commercial network environments is also required to verify the practical applicability of the proposed algorithms.
Finally, the proposed analytical framework and algorithms are expected to be applicable to other CS-based off-grid problems, such as CS-based channel estimation.}

\bibliographystyle{IEEEtran}
\bibliography{trans_ref}{}

\end{document}